\documentclass[12pt]{article} 
\usepackage{epsfig,amsmath,amssymb,subfigure}
\renewenvironment{itemize}
   {\begin{list}%
      {}%
      {\setlength{\topsep}{0pt}%
       \setlength{\partopsep}{0pt}%
       \setlength{\itemsep}{-4pt}%
       \setlength{\labelsep}{5pt}%
       \setlength{\itemindent}{0pt}%
      }%
   }%
   {\end{list}}%

\makeatletter
\renewcommand{\section}{\setcounter{equation}{0}\@startsection
  {section}%
  {1}%
  {0pt}%
  {-1\baselineskip}%
  {0.4\baselineskip}%
  {\bfseries\large}}%
\renewcommand{\subsection}{\@startsection
  {subsection}%
  {2}%
  {0pt}%
  {-0.75\baselineskip}%
  {0.2\baselineskip}%
  {\bfseries}}%
\renewcommand{\subsubsection}{\@startsection
  {subsubsection}%
  {3}%
  {0pt}%
  {-0.5\baselineskip}%
  {0.1\baselineskip}%
  {\sc}}%
\makeatother

\renewcommand{\theequation}{\thesection.\arabic{equation}}

\def\cA{{\cal A}}

\def\a{\alpha} 
\def\b{\beta}
\def\d{\delta}          

\def\ga{\gamma}         
\def\gm{\Gamma}
 
\def\la{\lambda}

\def\m{\mu}
\def\n{\nu}

\def\om{\omega}

\def\vslash{{v\mkern-9mu/}}

\def\Vslash{{V\mkern-11mu/}}

\def\prslash{{\partial\mkern-9mu/}}
\def\pslash{{p\mkern-8mu/}{\!}}

\def\prslash{{\partial\mkern-9mu/}}    
\def\qslash{{q\mkern-8mu/}{\!}}

\def\mutw{\mu_2}

\def\idp{\int\!\! \frac{d^4\!p}{(2\pi)^4}}

\def\idq{\int\!\! \frac{d^4\!q}{(2\pi)^4} \,\,}

\def\idqo{\int\!\! \frac{d^{2\omega}\!q}{(2\pi)^{2\omega}}}
\def\idx{\int\!\! d^4\!x}

\def\idxo{\int\!\! d^{2\omega}\!x}

\def\dotAstar{{\overset{~\,\circ}{\cal A}_\star}}
\def\dotAthree{\overset{~\,\circ}{\cal A}_{\star,3}}
\def\dotAfour{\overset{~\,\circ}{\cal A}_{\star,4}}
\def\dotAfive{\overset{~\,\circ}{\cal A}_{\star,5}}
\def\dotAsix{\overset{~\,\circ}{\cal A}_{\star,6}}

\def\ds{\displaystyle}
\def\scs{\scriptstyle}

\newcommand{\defpar}{t}

\newcommand{\lieg}{\mathfrak{g}}
\newcommand{\cB}{{\cal B}}
\newcommand{\cF}{{\cal F}}
\newcommand{\cP}{{\cal P}}
\newcommand{\cU}{{\cal U}}
\newcommand{\bea}{\begin{eqnarray}} 
\newcommand{\eea}{\end{eqnarray}}
\newcommand{\beann}{\begin{eqnarray*}} 
\newcommand{\eeann}{\end{eqnarray*}}
\newcommand{\ba}{\begin{array}}
\newcommand{\ea}{\end{array}}
  
\newcommand{\6}{\partial }

\newcommand{\f}[3]{{f_{#1#2}}^{#3}}

\newcommand{\ST}{\star}
\newcommand{\gh}{\mathrm{gh}}
\newcommand{\Dim}{\mathrm{dim}}
\newcommand{\LA}{{\5a}}
\newcommand{\LB}{{\5b}}
\newcommand{\LC}{{\5c}}

\begin{document}
\begin{titlepage}
\rightline{MPI-MIS-70/2003}
\vglue 45pt

\begin{center}
{\Large \bf Anomaly freedom in Seiberg-Witten\\[3pt] 
            noncommutative gauge theories}\\ 
\vskip 1.2 true cm {\rm Friedemann Brandt$^a$\footnote{E-mail:
fbrandt@mis.mpg.de}, C.P. Mart\'{\i}n}$^b$\footnote{E-mail:
carmelo@elbereth.fis.ucm.es} and F. Ruiz Ruiz$^c$\footnote{E-mail:
t63@aeneas.fis.ucm.es}
\vskip 0.3 true cm
{\it
$^a$Max-Planck-Institute for Mathematics in the Sciences,\\
Inselstra\ss e 22-26, D-04103 Leipzig, Germany}\\
\vskip 0.3 true cm
{\it $^{b,\,c}$ Departamento de F\'{\i}sica Te\'orica I, 
Facultad de Ciencias F\'{\i}sicas\\ 
Universidad Complutense de Madrid,
 28040 Madrid, Spain}\\

\vskip 1.2 true cm

{\leftskip=50pt \rightskip=50pt 
\noindent
We show that noncommutative gauge theories with arbitrary
compact gauge group defined by means of the Seiberg-Witten map have
the same one-loop anomalies as their commutative counterparts. This is
done in two steps. By explicitly calculating the
$\epsilon^{\m_1\m_2\m_3\m_4}$ part of the renormalized effective
action, we first find the would-be one-loop anomaly of the theory to
all orders in the noncommutativity parameter $\theta^{\m\n}$. And
secondly we isolate in the would-be anomaly radiative corrections
which are not BRS trivial. This gives as the only true anomaly
occurring in the theory the standard Bardeen anomaly of commutative
spacetime, which is set to zero by the usual anomaly cancellation
condition.\par }
\end{center}

\vspace{20pt}
\vfill
\end{titlepage}


\setcounter{page}{2}
\section{Introduction}

It is a well known fact that not all relevant gauge groups in particle
physics are consistent with the Moyal product of noncommutative field
theory. An example of this is provided by the Moyal product
$A_\mu(x)\star A_\nu(x)$ of two $SU(N)$ Lie algebra valued gauge
fields $A_\mu(x)$ and $A_\nu(x)$. It is clear that such product does
not lie in the $SU(N)$ Lie algebra but in a representation of its
enveloping algebra, so $A_\mu(x)$ can not be regarded as a truly
noncommutative $SU(N)$ gauge field. This makes it difficult to
formulate, even classically, noncommutative extensions of some
physically relevant gauge theories like {\it e.g.} the standard
model. A way to circumvent this problem
\cite{Madore:2000en,Jurco:2000ja} is to build noncommutative gauge and
matter fields from ordinary ones by means of the Seiberg-Witten map
\cite{Seiberg:1999vs}. Using this approach, classical noncommutative
gauge theories have been constructed for arbitrary compact groups
\cite{Madore:2000en,Jurco:2000ja,Jurco:2001my,Jurco:2001rq} and
noncommutative gauge theories with $SU(5)$ and $SO(10)$ gauge groups
have been constructed in ref.~\cite{Aschieri:2002mc}. Furthermore, a
noncommutative standard model has been formulated in
ref. \cite{Calmet:2001na} and some of its phenomenological
consequences have been explored in a number of papers
\cite{Carroll:2001ws,Carlson:2001sw,Behr:2002wx,Carlson:2002zb,Schupp:2002up,Minkowski:2003jg}. Many
of these noncommutative gauge theories, among them the noncommutative
standard model, involve chiral fermions, so the corresponding
classical gauge symmetry may be broken by quantum corrections. In
other words, an anomaly may occur and the resulting quantum theory may
then become inconsistent. To study the consistency of quantum
noncommutative gauge theories defined by means of the Seiberg-Witten
map, it is therefore necessary to study whether new types of anomalies
occur --{\it i.e.}  anomalies which do not appear in ordinary
commutative spacetime and hence that may require additional anomaly
cancellation conditions.

In refs. \cite{Barnich:ve,Barnich:mt,Barnich:2000zw} it has been shown
that for Yang-Mills type gauge theories with arbitrary semisimple
gauge groups the only nontrivial solution to the anomaly consistency
condition is the usual Bardeen anomaly, regardless of whether or not
the theory is Lorentz invariant or renormalizable by power
counting. This result readily applies to gauge noncommutative field
theories constructed by means of the Witten-Seiberg map, since, as far
as these matters are concerned, the presence of a noncommutative
matrix parameter $\theta^{\m\n}$ with mass dimension $-2$ only
precludes Lorentz invariance and power-counting
renormalizability. Thus, for noncommutative gauge theories with
semisimple gauge groups, there are no $\theta^{\m\n}\!\!$-dependent
anomalies and any $\theta^{\m\n}\!\!$-dependent breaking of the BRS
identity, being cohomologically trivial, can be set to zero by adding
appropriate counterterms to the effective action. Note that the
addition of these $\theta^{\m\n}\!\!$-dependent counterterms to the
effective action makes sense within the framework of effective field
theory, but this agrees with the observation that noncommutative field
theories defined by means of the Seiberg-Witten map should be
considered as effective field theories
\cite{Aschieri:2002mc,Wulkenhaar:2001sq}. All the above implies that
no anomalous $\theta^{\m\n}\!\!$-dependent terms should occur in the
Green functions of noncommutative theories with semisimple gauge
groups, a fact that has been proved to hold true at order one in
$\theta^{\m\n}$ for the three-point function of the gauge field and a
simple gauge group by explicit computation of the appropriate Feynman
diagrams \cite{Martin:2002nr}.

The situation is very different if the gauge group is not
semisimple. In this case, the consistency condition for gauge
anomalies has other nontrivial solutions besides Bardeen's anomaly. 
In particular, in four dimensions and if the
gauge group is $G\times U(1)_{\rm Y}$, with $G$ a semisimple gauge
group, the additional nontrivial solutions are of the form
\begin{equation}
 \idx ~c\>{\rm I}_{\rm inv}[f_{\mu\nu},G_{\m\n}]~.
\label{candidates}
\end{equation}
Here matter fields have been integrated out, $c$ is the $U(1)_{\rm Y}$
ghost field and ${\rm I}_{\rm inv}[f_{\m\n},G_{\m\n}]$ is a gauge
invariant function of the $U(1)_{\rm Y}$ field strength $f_{\m\n}$,
the $G$ field strength $G_{\m\n}$ and their covariant derivatives.
Note that there are infinitely many candidate anomalies of this type
since neither power counting nor Lorentz invariance are available to
reduce the number of invariants ${\rm I}_{\rm
inv}[f_{\m\n},G_{\m\n}]$. Furthermore, when the gauge group contains
more than one abelian factor, there are additional candidate anomalies
of yet another type \cite{Barnich:ve,Barnich:mt,Barnich:2000zw}.  The
purpose of this paper is to investigate whether anomalies of these
types occur in noncommutative gauge theories with nonsemisimple gauge
groups defined through the Seiberg-Witten map. This is not a trivial
question and has far reaching implications. Indeed, did solutions of
type (\ref{candidates}) occur in perturbation theory, the
corresponding quantum gauge theory would be anomalous, the anomaly
being $\theta^{\m\n}\!\!$-dependent. To remove the resulting anomaly
and render the quantum theory consistent, one would then have to
impose constraints on the fermions hypercharges. A conspicuous
instance of a model with such a gauge group for which this point
should be cleared is the noncommutative standard model
\cite{Calmet:2001na}.

In this paper we will prove that, for a noncommutative field theory
with arbitrary compact gauge group defined by means of the
Seiberg-Witten map, the only anomaly that occurs at one loop (hence,
to all orders in perturbation theory, if one assumes the existence of
a nonrenormalization theorem for the anomaly) is the usual Bardeen
anomaly. The paper is organized as follows. In Section 2 we fix the
notation, define the chiral BRS transformations and use the
Seiberg-Witten map to classically define the noncommutative
model. Section 3 uses dimensional regularization to explicitly compute
the $\epsilon^{\m_1\m_2\m_3\m_4}$ part of the renormalized effective
action. This yields a complicated power series in the noncommutativity
parameter $\theta^{\m\n},$ of which the term of order zero is the
usual Bardeen anomaly of commutative field theory. In Section 4 we
show that all terms in this series of order one or higher in
$\theta^{\m\n}$ are cohomologically trivial with respect to the chiral
BRS operator and find the counterterm that removes them from the
renormalized effective action. Section 5 contains our conclusions. We
postpone to two appendices some very technical points of our
arguments. Let us emphasize that in this paper we will only discuss
gauge anomalies --see refs. \cite{Banerjee:2001un,Banerjee:2003ce} for
related work on the rigid axial anomaly.

\section{The model, notation and conventions}\label{sec2}

Let us consider a compact nonsemisimple gauge group $G = G_1 \times
\cdots \times G_N$, with $G_i$ a simple compact group if
$\,i=1,\ldots,s\,$ and an abelian group if $\,i=s+1,\ldots,N$. We may
assume without loss of generality that the abelian factors come with
irreducible representations, which of course are one-dimensional. Let
us denote by $\psi_{i_1\cdots i_s}$ a Dirac field on ordinary
Minkowski spacetime carrying an arbitrary unitary irreducible
representation of the Lie algebra of $G$. Since the abelian factors
come with one-dimensional representations, the indices in the Dirac
field $\psi_{i_1\cdots i_s}$ correspond to the simple factors. From
now on we will collectively denote the ``simple'' indices $(i_1\cdots
i_s)$ by the multi-index $I$. The corresponding vector potential
$v_\m$ on Minkowski spacetime in the representation of the Lie
algebra carried by $\psi_I$ will have the following decomposition in
terms of the gauge fields $a_\m^k$ and $a_\mu^l$ associated to the
factors of the group $G$
\begin{equation*}
    v_\m = \sum_{k=1}^s g_k\, (a_\mu^k)^a\,(T^k)^a 
         + \sum_{l=s+1}^N g_l\,a_\mu^l\,T^l~.
\end{equation*}
Here $g_k$ and $g_l$ are the coupling constants and $\{(T^k)^a,T^l\}$,
with $a=1,\ldots,{\rm dim}\,G_k$ for every $k=1,\ldots,s$ and
$l=s+1,\ldots,N$, stand for the generators of the $G$ Lie algebra in
the unitary irreducible representation under consideration. As usual,
a sum over $a$ is understood. The matrix elements $IJ$ of these
generators are always of the form
\begin{equation*}
\begin{array}{c}
    (T^k)^a_{IJ} = \d_{i_1 j_1} \cdots (T^k)^a_{i_k j_k} 
        \cdots \d_{i_s j_s} \\[6pt]
    T^l_{IJ} = \d_{i_1 j_1} \cdots \d_{i_s j_s} Y^l ~,
\end{array}     
\end{equation*}
where $(T^k)^a_{i_k j_k}$ are the matrix elements of the generator
$(T^k)^a$ of the Lie algebra of the factor $G_k$ in some given
irreducible representation. Given any two generators $(T^k)^a_{IJ}$
and $(T^{k'})^{a'}_{IJ}$ as above we define the trace operation
$\mathbf{Tr}$ as
\begin{align*}
   \mathbf{Tr}\>(T^k)^a\,(T^{k'})^{a'} & = 
     (T^k)^a_{IJ}\,(T^{k'})^{a'}_{JI} \\[3pt]
   & = \d_{i_1 j_1} \cdots (T^k)^a_{i_k j_k} \cdots \d_{i_s j_s} 
           \d_{j_1 i_1} \cdots (T^{k'})^{a'}_{j_{k'} i_{k'}} 
           \cdots \d_{j_s i_s} ~.
\end{align*}
The ghost field $\lambda$ associated to
$v_\m$, also in the representation furnished by $\psi_I$, is
\begin{equation*}
    \lambda = \sum_{k=1}^s g_k\,(\lambda^k)^a\,(T^k)^a 
         + \sum_{l=s+1}^N g_l\,\lambda^l\,T^l~,
\end{equation*}
with $(\lambda^k)^a$ and $\lambda^l$ being the ghost fields for the
factors in $G$. Now we consider the theory that arises from chirally
coupling, say left-handedly, the fermion field $\psi_I$ to the gauge
field $v_\m$. The fermionic part of the corresponding classical
action reads
\begin{equation}
    S^{\rm fermion} = \idx\,\bar{\psi}_I\, i\hat{D}(v)_{IJ}\,\psi_J ~,
\label{classact}
\end{equation}
with $\bar{\psi}_I=\psi_I^{\dagger}\ga^0$ and
\begin{equation*}
    \hat{D}(v)_{IJ}\,\psi_J = \d_{IJ}\,\prslash\psi_J 
        + \vslash_{IJ}\,{\rm P}_{-}\psi_J~.
\end{equation*}
Here ${\rm P}_{-}$ is the left-handed chiral projector, given by
\begin{equation*}
    \qquad  
    {\rm P}_\pm = \frac{1}{2}\>(1\pm \ga_{5}) \qquad
    \ga_5 = -i\ga^0\ga^1 \ga^2 \ga^3 ~,
\end{equation*}
 the gamma matrices $\ga^\m$ being defined by
$\{\ga^\m,\ga^\n\}=2\eta^{\mu\nu}$ and the convention for the
Minkowski metric $\eta_{\m\n}$ being $\eta_{\m\n}={\rm
diag}\,(+,-,-,-)$.  This action is invariant under the chiral BRS
transformations
\begin{equation}
    s v_{\mu}=\partial_{\mu}\lambda + [v_\m,\lambda] \qquad 
    s \psi= -\lambda {\rm P}_{-}\psi \qquad 
    s \bar{\psi} = \bar{\psi}\lambda {\rm P}_+ \qquad
    s\lambda=-\lambda\,\lambda ~.
\label{BRStrans}
\end{equation}
As usual, the BRS operator $s$ commutes with $\partial_\m$, satisfies
the anti-Leibniz rule and is nilpotent, {\it i.e.} $s^2=0$.

To construct the noncommutative extension of the ordinary gauge theory
defined by the classical action $S^{\rm fermion}\!$, we use the
formalism developed in refs. \cite{Madore:2000en, Jurco:2000ja,
Jurco:2001my, Jurco:2001rq}. To this end, we first define the
noncommutative gauge field $V_\m$, the noncommutative spinor field
$\Psi_I$ and the noncommutative ghost field $\Lambda$ in terms of
their ordinary counterparts $v_\m$, $\psi_I$ and $\lambda$ by using
the Seiberg-Witten map \cite{Seiberg:1999vs}. This is done as
follows. The fields $V_\m= V_\m\, [v;\theta]$, $\Psi_I=\Psi_I\,
[\psi,v;\theta]$ and $\Lambda= \Lambda\,[\lambda, v;\theta]$ are
formal power series in $\theta^{\m\n}\!$, with coefficients depending
on the ordinary fields and their derivatives, that take values in the
representation of the enveloping algebra of the Lie algebra of the
group $G$ furnished by the ordinary Dirac field $\psi_I$ and solve
the Seiberg-Witten equations
\begin{equation} 
    s_\star V_\m = s V_\mu \qquad 
    s_\star \Psi = s \Psi \qquad
    s_\star \Lambda = s\Lambda
\label{SWequations}
\end{equation}
subject to the boundary conditions
\begin{equation}
    V_\m\, [v;\theta\!=\!0] = v_\mu \qquad
    \Psi_I\,[v,\psi;\theta\!=\!0] = \psi_I \qquad
    \Lambda\,[\lambda, v;\theta\!=\!0] = \lambda ~.
\label{Boundary}
\end{equation}
In eq. (\ref{SWequations}) $s$ is the ordinary BRS operator of
eq. (\ref{BRStrans}), while $s_\star$ denotes the noncommutative BRS
chiral operator, whose action on the noncommutative fields is given by
\begin{equation}
   s_\star V_\m = \partial_\m \Lambda + [V_\m,\Lambda]_{\star} \qquad
   s_\star \Psi = - \Lambda \star {\rm P}_{-}\Psi\qquad
   s_\star \Lambda = -\,\Lambda \star \Lambda ~.
\label{NCBRStrans}
\end{equation}
The commutator $[f,g]_\star$ stands for
\begin{equation*}
   [f,g]_\star= f \star g - g \star f~,
\end{equation*}
with $\,f\star g\,$ the Moyal product of functions on Minkowski
spacetime, defined for arbitrary $f$ and $g$ by
\begin{displaymath}
   (f\star g)(x) = \idp \! \idq e^{-i(p+q)x}\,
      e^{-\frac{i}{2}\,\theta^{\a\b}p_\a q_\b}\,
     \tilde{f}(p) \> \tilde{q}(q)~,
\end{displaymath}
$\tilde{f}(p)$ and $\tilde{g}(q)$ being the Fourier transforms of $f$
and $g$. For the noncommutative field $\Psi_I$ we further demand it to
be linear in $\psi_I$, so that
\begin{equation}
   \Psi_{\a I} =\Big( \d_{IJ}\,\d_{\a\b} 
      + M[v,\partial,\ga,\ga_5;\theta]_{\a\b\;IJ}\Big) \psi_{\b J}~,
\label{ncfermion}
\end{equation}
where $\a$ and $\b$ are Dirac indices. Note that, in accordance with
the boundary condition for $\Psi_I\,[\psi,v;\theta]$, the differential
operator $M[v,\partial,\ga,\ga_5;\theta]_{\a\b\,IJ}$ vanishes at
$\theta=0$. Taking $\Psi_i$ linear in $\psi_I$, as in
eq. (\ref{ncfermion}), is always possible \cite{Cerchiai:2002ss} and is the
natural choice within the framework of noncommutative
geometry~\cite{Wulkenhaar:2001sq}. Once the noncommutative fields have
been defined, one considers the following noncommutative classical
action
\begin{equation}
    S^{\rm fermion}_{\rm nc} = \idx\,\bar{\Psi}_I \star 
     i\hat{D}(V)_{IJ} \Psi_J~,
\label{ncact}
\end{equation}
where $\,\bar{\Psi}_I=\bar{\Psi}_I^\dagger\ga^0$ and
\begin{equation*}
   \hat{D}(V)_{IJ} \Psi_J = \d_{IJ} \prslash \Psi_J 
         + \Vslash_{IJ} \star {\rm P}_{-}\Psi_J ~.
\end{equation*}
We stress that the noncommutative fields are functions of ordinary
fields as given by the Seiberg-Witten map and hence the noncommutative
action is also a functional of these. Furthermore, the noncommutative action
$S^{\rm fermion}_{\rm nc}$ is invariant under the ordinary chiral BRS
transformations in eq. (\ref{BRStrans}) since, by definition of the
Seiberg-Witten map,
\begin{equation*}
   sS^{\rm fermion}_{\rm nc} = s_\star S^{\rm fermion}_{\rm nc}
\end{equation*}
and, by construction, 
\begin{equation*}
    s_\star S^{\rm fermion}_{\rm nc}=0~.
\end{equation*}

The effective action $\Gamma[v;\theta]$ of the noncommutative theory
is formally defined by
\begin{equation}
\begin{array}{c}
   {\ds \Gamma[v;\theta] = -i \ln Z[v;\theta] }\\[6pt]
   {\ds Z[v;\theta] = {\cal N} {\ds \int [d\bar{\psi}]\>[d\psi] ~ 
         \exp\,\Big(i\,S_{\rm nc}^{\rm fermion}\Big) ~,}} 
\end{array}
\label{effectiveaction}
\end{equation}
with ${\cal N}$ a normalization constant chosen so that
$Z[v\!=\!0;\theta]=1$, {\it i.e.}
\begin{equation*}
   {\cal N}^{-1} = \int  [d\bar{\psi}]\>[d\psi]~ \exp \left( 
      \idx\>\bar{\psi}\,i\prslash\psi\right)~,
\end{equation*}
and $[d\bar{\psi}]\>[d\psi]$ the measure for ordinary fermion
fields. Also formally, the invariance of $S^{\rm fermion}_{\rm nc}$
under $s$ leads to the invariance of $\Gamma[v;\theta]$ under ordinary
gauge transformations of $v_\m$. The problem is that all this is
formal since defining the effective action requires
renormalization. The question that should really be addressed is
whether it is possible to define a renormalized effective action
$\Gamma^{\rm ren}[v;\theta]$ invariant under $s$.  Were this the case,
the theory would be anomaly free. In this paper we provide an answer
in the negative and show that the anomaly has the same form as for the
ordinary, {\it i.e.} commutative theory.

\section{Form of the noncommutative anomaly}

In this section we use dimensional regularization to define a
renormalized effective action $\Gamma^{\rm ren}[v;\theta]$ and find a
closed expression for the anomaly $s\Gamma^{\rm ren}[v;\theta]$ in
terms of the noncommutative fields $V_\m$ and $\Lambda$. To
dimensionally regularize the theory, we consider the action
\begin{equation}
    S^{\rm reg}_{\rm nc} = \idxo\,\bar{\Psi}_I\star 
       i\Big(\d_{IJ}\, \prslash\Psi_{J} 
     + \bar{\ga}^\m V_{\mu\,IJ}\star {\rm P}_{-}\Psi_J\,\Big)~,
\label{regact}
\end{equation}
first introduced in the context of noncommutative gauge theories in
ref. \cite{Martin:2002nr} for $U(N)$ theories and theories with simple
groups. Here we use dimensional regularization {\it \'a la}
Breitenlohner and Maison \cite{Breitenlohner:hr}. We will use the
notation in that reference, in which 4-dimensional objects are denoted
with bars $(\bar{g}^\m_{~\m}=4)$ and evanescent or
$(2\omega\!-\!4)\!$-dimensional quantities are denoted with hats
$(\hat{g}^\m_{~\m}=2\omega-4)$. The dimensionally regularized
partition function $Z^{\rm reg}[v;\theta]$ is defined as the sum of
the dimensionally regularized Feynman diagrams generated by the path
integral
\begin{equation} 
   Z^{\rm reg}[v;\theta] = {\cal N}\int [d\bar{\psi}]\,[d\psi]~
          \exp \Big( i\,S_{\rm nc}^{\rm reg}\Big) ~.
\label{regpartfunc}
\end{equation}

In the regularized partition function we perform the change of
variables $ \psi_{\b J},\bar{\psi}_{\b J}\to \Psi_{\a
I},\bar{\Psi}_{\a I}$, with
\begin{equation}
\begin{array}{c}
   \Psi_{\a I} = \big( \d_{IJ}\,\d_{\a\b} + 
     M[v,\partial,\ga,\ga_5;\theta]_{\a\b\;IJ}\big)\, \psi_{\b J} \\[6pt]
   \bar{\Psi}_{\a I} = \big(  \d_{IJ}\,\d_{\a\b} 
      + \bar{M}[v,\partial,\ga,\ga_5;\theta]_{\a\b\;IJ}\big)\,
                                                 \bar{\psi}_{\b J} \\[6pt]
   [d\bar{\psi}]\,[d\psi] = \det\big(\mathbb{I} + \bar{M}\big)\,
      \det\big(\mathbb{I} + M\big)\,[d\bar{\Psi}]\,[d\Psi]~,\\[6pt]
\end{array}
\label{change}
\end{equation}
where the determinants are defined by their diagrammatic expansion in
dimensional regularization in powers of $\theta$. Now, in dimensional
regularization we have
\begin{equation}
    \det\big(\mathbb{I} + M\big) = \det\big(\mathbb{I} + \bar{M}\big)
         = 1~.
\label{trivialdet}
\end{equation}
To see this, take {\it e. g.} the determinant $\det\big(\mathbb{I} +
M\big)$ and write it as the partition function
\begin{equation*}
    \det\big(\mathbb{I} + M\big) = \int [d\bar{\psi}]\,[d\psi]~ 
         e^{iS[M]}
\end{equation*}
of a fermion theory with classical action
\begin{equation*}
     S[M] = \idxo~ \bar{\psi}\,\big( \mathbb{I} 
        + M \big)\,\psi~.
\end{equation*}
The propagator of such a theory is the identity and the interaction
vertices come from the operator $M[v,\partial,\ga,\ga_5;\theta]$, so the
Feynman integrals that enter the diagrammatic expansion of
$\det\big(\mathbb{I} + M\big)$ are of the form
\begin{equation*}
    \idqo q_{\m_1}\cdots q_{\m_n}~.
\end{equation*}
Since this integral vanishes in dimensional regularization,
eq. (\ref{trivialdet}) holds and the change of variables (\ref{change})
gives for the path integral in (\ref{regpartfunc})
\begin{equation}
   Z^{\rm reg}[v;\theta] = {\cal N} \int [d\bar{\Psi}]\,[d\Psi] ~ 
       \exp \Big( i S_{\rm nc}^{\rm reg}[\bar{\Psi},\Psi,V_\m]\Big) ~.
\label{partitionfunction}
\end{equation}
Hence $Z^{\rm reg}[v;\theta]$  is a functional of $V_\m$,
and so is the regularized effective action 
\begin{equation}
   \Gamma^{\rm reg}[v;\theta] = -i \ln Z[v;\theta]^{\rm reg} 
        = \Gamma^{\rm reg}[V] ~.
\label{reg-effec-action}
\end{equation}
In other words, the regularized effective action depends on $v_\m$
through $V_\m$. 

Eq.~(\ref{reg-effec-action}) for $\Gamma^{\rm reg}[V]$ is to be
understood in a diagrammatic sense as the generating functional of 1PI
Green functions for the field $V_\m$. That is to say,
\begin{equation}
  i\Gamma^{\rm reg}[V] = \sum_{n=1}^\infty~\frac{1}{n!}\>
    \idxo_1 \ldots \idxo_n~  V_{\m_1 I_1J_1}(x_1) \> \ldots 
       V_{\m_n I_nJ_n}(x_n)\> \Gamma^{\m_1\ldots\m_n}_{I_1J_1 \ldots 
          I_nJ_n}(x_1,\ldots,x_n) ~,
\label{reg-effec-action-expansion}
\end{equation}
with 
\begin{equation}
  \Gamma^{\m_1\ldots\m_n}_{I_1J_1\,\ldots\, I_nJ_n}
     (x_1,\ldots,x_n ) = {\langle {\cal J}^{\m_1}_{I_1J_1}(x_1)\>
        \ldots {\cal J}^{\m_n}_{I_nJ_n}(x_n) \rangle}_0^{\rm conn}
\label{Greenfunction}
\end{equation}
and
\begin{equation}
   {\cal J}^{\m_i}_{I_iJ_i}(x_i) = (\Psi_{\b_i J_i} \star 
       \bar{\Psi}_{\a_i I_i})(x_i)\>
           (\bar{\ga}^{\m_i}\ga_5)_{\a_i\b_i} ~.
\label{currents}
\end{equation}
Here the symbol ${\langle {\cal O}\rangle}_0^{\rm conn}$ stands for
the connected component of the correlation function ${\langle {\cal
O}\rangle}_0$ defined by
\begin{equation}
   {\langle{\cal O}\rangle}_0 = \int [d\bar{\Psi}]\,[d\Psi] ~{\cal O}~   
     {\rm exp}\,\bigg( i\! \idxo \>\bar{\Psi}_I \prslash \Psi_I\bigg) ~.
\label{freemeasure}
\end{equation}
Note that eqs. (\ref{Greenfunction}) and (\ref{freemeasure}) define
$\,\Gamma^{\m_1\ldots\m_n}_{I_1J_1 \ldots I_nJ_n} (x_1,\ldots,x_n)\,$
as the result of applying Wick's theorem to $\,{\cal
J}^{\m_1}_{I_1J_1}(x_1) \ldots {\cal J}^{\m_n}_{I_nJ_n}(x_n)\,$ with
regard to the contraction
\begin{equation}
    {\langle\Psi_{\b J}(y)\,\bar{\Psi}_{\a I}(x)\rangle}_0 = \d_{JI} 
        \idqo ~e^{-iq(y-x)}~\frac{i \qslash_{\b\a}}{q^2+i0^+}~.          
\label{contraction}
\end{equation} 
It is not difficult to see that in eq.  (\ref{Greenfunction}) there
are $(n-1)!$ different contractions and that, upon combination with
the $V'{\rm s}$ in eq. (\ref{reg-effec-action-expansion}), they all
yield the same contribution. The regularized effective action then
takes the form
\begin{equation}
  i\Gamma^{\rm reg}[V] = - \sum_{n=1}^\infty~\frac{(-1)^n}{n}\>
  \idxo_1  \ldots \idxo_n~ \mathbf{Tr}\, 
        [V_{\m_1}(x_1) \ldots V_{\m_n}(x_n)]\> 
             \Gamma^{\m_1\ldots\m_n}(x_1,\ldots,x_n) ~,
\label{firstresult}
\end{equation}
where
\begin{equation*}
  {\mathbf{Tr}\, \big[V_{\m_1}(x_1)\ldots V_{\m_n}(x_n)\big] 
     = V_{\m_1 I_1 I_2}(x_1)\> V_{\mu_2 I_2 I_3}(x_2) \ldots 
        V_{\m_{n-1} I_{n-1} I_n}}(x_{n-1})\>
            V_{\m_n\,I_n I_1}(x_n) ~,
\end{equation*}
the 1PI Green function $ \gm_{\m_1\ldots\m_n}(x_1,\ldots,x_n)$ reads
\begin{equation}
\begin{array}{l}
   {\ds \gm_{\m_1\ldots\m_n}(x_1,\ldots,x_n) = i^n\! \int \prod_{i=1}^n\,
     \frac{d^{2\omega}\!p_i}{(2\pi)^{2\omega}}~ 
       (2\pi)^{2\omega}\delta(p_1+\ldots+p_n) ~
          e^{\, i\!\sum\limits_{i=1}^{n} p_i x_i} ~  e^{ -\frac{i}{2}\!\! 
             \sum\limits_{1\leq i<j<n}\theta^{\a\b} p_{i\a}p_{j\b}} }\\[12pt]
\phantom{\Gamma^{\m_1\ldots}~}
   {\ds \times \idqo~
      \frac{\mathrm{tr}\, \big[\,(\qslash+\pslash_1)\, 
          \bar{\ga}^{\m_1}{\rm P}_{-}\, \qslash\, \bar{\ga}^{\m_2}\, 
            {\rm P}_{-}\,(\qslash-\pslash_2)\,\ldots\, 
               \big(\qslash-\sum_{i=2}^{n-1}\pslash_i\big)\,
                  \bar{\ga}^{\m_n}\,{\rm P}_{-}\,\big]}
                     {(q+p_1)^2\,q^2\,(q-p_2)^2\cdots 
                        (q-\sum_{i=2}^{n-1}p_i)^2} } 
\end{array}
\label{gammanint}
\end{equation}
and the symbol $\mathrm{tr}$ denotes trace over Dirac matrices. For
completeness we present very briefly an alternative derivation of
(\ref{firstresult}). Integrating over $[d\bar{\Psi}]$ and $[d\Psi]$ in
eq. (\ref{partitionfunction}) and using eq. (\ref{reg-effec-action}),
we obtain
\begin{equation}
   i \Gamma^{\rm reg}[V] = {\rm Tr}\,\ln \left[ 1 +  (\prslash)^{-1}
      \bar{\ga}^\m V_\m {\rm P}_{-}\star \right] 
   =  - \sum_{n=1}^n \>\frac{(-1)^n}{n}\> {\rm Tr}\, 
      {\left[ (\prslash)^{-1}\bar{\ga}^\m V_\m {\rm P}_{-} 
        \star\right]}^n~, 
\label{secondresult}
\end{equation}
where ${\rm Tr}$ is to be interpreted as ${\ds \idxo}$ for the
continuous indices of the operator on which ${\rm Tr}$ acts and
$(\prslash)^{-1}$ has matrix elements $ \langle
y|(\prslash)^{-1}|x\rangle$ given by the right-hand-side of
eq. (\ref{contraction}).  Clearly the right-hand side of
eq. (\ref{secondresult}) has a neat diagrammatic representation which
readily leads to eq. (\ref{firstresult}).

We stress the fact that the noncommutative field $V_\m(x)$ in
eqs. (\ref{firstresult}) and (\ref{secondresult}) is a mere spectator
in the sense that these equations hold whatever the algebra on which
$V_\m(x)$ takes values be, provided the operation $\mathbf{Tr}$ make
sense. Eqs. (\ref{firstresult}) and (\ref{secondresult}) are thus
valid for noncommutative $U(N),$ simple, semisimple and non-semisimple
gauge groups. One then expects that for nonsemisimple gauge groups a
renormalized effective action $\Gamma^{\rm ren}[V]$ can be defined so
that the noncommutative gauge anomaly has the same form as for
noncommutative $U(N)$ group, {\it i.e.} such that
\begin{equation}
   s_\star\Gamma^{\rm ren}[V] = {\cal A}_\star ~,
\label{nonsemianom-1} 
\end{equation}
with
\begin{equation}
   {\cal A}_\star = - \frac{i}{24\pi^2}
         \idx~ \epsilon^{\m_1\m_2\m_3\m_4}\> \mathbf{Tr}\>
              \Lambda \star \partial_{\m_1}\Big( 
                 V_{\m_2}\star\partial_{\m_3} V_{\mu_4} 
       + \frac{1}{2}\> V_{\m_2}\star V_{\m_3}\star V_{\m_4}\Big)~.
\label{nonsemianom-2}
\end{equation}
In the remainder of the section we prove that is indeed so. 

To demonstrate eqs. (\ref{nonsemianom-1}) and (\ref{nonsemianom-2}) we
proceed as follows. Since the integral over $d^{2\omega}q$ in
eq. (\ref{firstresult}) does not involve any nonplanar factor
$e^{iq_\a\theta^{\a\b} p_{i\b}}$, the effective action in
eq. (\ref{firstresult}) is given by a sum over dimensionally
regularized planar diagrams. Hence, the Quantum action principle
\cite{Breitenlohner:hr} holds for this effective action and the
following equation is valid
\begin{equation}
     s_\star \Gamma^{\rm reg}[V] = \hat{\Delta} \cdot 
       \Gamma^{\rm reg}[V] ~.
\label{regbreaking}
\end{equation} 
Here $\hat{\Delta}\!\cdot\!\Gamma^{\rm reg}[V]$ is the insertion in
$\Gamma^{\rm reg}[V]$ of the evanescent operator $\hat{\Delta}$
defined by
\begin{equation*}
  \hat{\Delta} = s_\star S_{\rm nc}^{\rm reg} = \idxo \> \left[ 
    \bar{\Psi}_I\star \Lambda_{IJ}\star i\hat{\prslash} {\rm P}_{+}\Psi_J 
      - \bar{\Psi}_I \star i\hat{\prslash}\, 
           (\Lambda_{IJ}\star{\rm P}_{-}\Psi_J) \right]~.
\end{equation*}
Substituting this in eq. (\ref{regbreaking}), we obtain for its
right-hand side
\begin{equation}
\begin{array}{l}
   {\ds \hat{\Delta} \cdot \Gamma^{\rm reg}[V] = -\sum_{n=1}^{\infty}\,
       (-1)^n \idxo \idxo_1 \ldots\idxo_n }\\[9pt]
\phantom{\ds \hat{\Delta} \cdot \Gamma^{\rm reg}[V] = \>} 
   {\ds \times\,\mathbf{Tr}\, \big[\Lambda(x)\, V_{\m_1}(x_1) \ldots 
      V_{\m_n}(x_n) \big] \> 
       \Gamma^{\m_1\ldots\m_n}(x,x_1,\ldots,x_n \mid\hat{\Delta})~, }
\end{array}
\label{insertone}
\end{equation}
where 
\begin{equation}
\begin{array}{l}
   {\ds \Gamma^{\m_1\ldots\m_n}(x,x_1,\ldots,x_n\mid \hat{\Delta})
      = i^{n+1} \int \frac{d^{2\omega}\!p}{(2\pi)^{2\omega}}  
         \int \prod_{i=1}^{n}\> \frac{d^{2\omega}\!p_i}{(2\pi)^{2\omega}}~
             e^{ i\big( px+\sum\limits_{i=1}^{n} p_i x_i\big)} ~}\\[12pt] 
\phantom{\Gamma^{\m_1\ldots\m_n}~}
   {\ds \times\, e^{ -\frac{i}{2}\!\! \sum\limits_{1\leq i<j<n} 
         \theta^{\a\b} p_{i\a}p_{j\b}}\> (2\pi)^{2\omega}\,
           \delta(p+p_1+\cdots p_n) \> \Gamma^{\m_1\ldots\mu_n} 
               (p,p_1,\ldots,p_n\mid \hat{\Delta}) ~,}\\[12pt]
\end{array}
\label{insertiontwo}
\end{equation}
and the 1PI Green function $\Gamma^{\m_1\ldots\mu_n}
(p,p_1,\ldots,p_n\mid \hat{\Delta})$ with the insertion reads
\begin{equation}
\begin{array}{l}
    {\ds \Gamma^{\m_1\ldots\mu_n}(p,p_1,\ldots,p_n\mid \hat{\Delta}) 
        = \idqo~ \frac{1}{q^2\,(q-p_1)^2\,(q-p_1-p_2)^2\,\ldots\, 
          {(q-\sum_{i=1}^{n} p_i)}^2} }\\[9pt]
\phantom{\Gamma~~}
    {\ds \times\,{\rm tr}\,\big[ \hat{\qslash}\,{\rm P}_{+} - 
       \big(\hat{\qslash}-\!\sum_{i=1}^{n} \hat{\pslash}_i\big)\, 
         {\rm P}_{-} \big] \, \qslash\, \bar{\ga}^{\m_1}{\rm P}_{-}\, 
           (\qslash-\pslash_1)\, \bar{\ga}^{\m_2}\,{\rm P}_{-}\,
              (\qslash-\pslash_1-\pslash_2)\, \ldots\, 
                \bar{\ga}^{\m_n} {\rm P}_{-}\,
                   \big(\qslash-\!\sum_{i=1}^{n}\pslash_i\big)\,. }
\end{array}
\label{insertionintegral}
\end{equation}
As before, $\mathrm{tr}$ denotes trace over Dirac matrices. For $n\geq
5$ the integral in eq. (\ref{insertionintegral}) is UV finite by power
counting at $2\omega=4$. Hence,
\begin{equation}
   \Gamma^{\m_1\ldots\mu_n}(p,p_1,\ldots,p_n\mid \hat{\Delta}) = 
        O(\varepsilon) \qquad n\geq 5~,
\label{integralone}
\end{equation}
where $\varepsilon=\omega-2$. As concerns $n\leq 4$, using the results
in Appendix A, it is straightforward to compute the contribution
$\Gamma^{\m_1\ldots\mu_n}_{\rm eps}(p,p_1,\ldots,p_n\mid
\hat{\Delta})$ to $ \Gamma^{\m_1\ldots\mu_n}(p,p_1,\ldots,p_n\mid
\hat{\Delta})$ involving $\epsilon^{\m_1\m_2\m_3\m_4}$. After some
calculations, we obtain
\begin{equation}
\begin{array}{c}
   {\ds \Gamma^{\m_1}_{\rm eps}(p,p_1\mid \hat{\Delta}) = 0 }\\[9pt]
   {\ds \Gamma^{\m_1\m_2}_{\rm eps}(p,p_1,p_2\mid \hat{\Delta}) = 
      \frac{1}{24\pi^2}\> \epsilon^{\rho\m_1\sigma\mutw} \, 
        p_{1\rho}\,p_{2\sigma} + O (\varepsilon) }\\[9pt]
   {\ds \Gamma^{\m_1\m_2\m_3}_{\rm eps}(p,p_1,p_2,p_3\mid
     \hat{\Delta}) = -\frac{1}{2} \> \frac{1}{24\pi^2}\> 
      \epsilon^{\rho\m_1\m_2\m_3}\,(p_1+p_2+p_3)_{\rho} 
        + O(\varepsilon) }\\[9pt] 
   {\ds \Gamma^{\m_1\m_2\m_3\m_4}_{\rm eps}(p,p_1,p_2,p_3,p_4\mid
      \hat{\Delta}) = O(\varepsilon) ~.}\\[9pt]
\end{array}
\label{nontrivialint}
\end{equation}
Substituting eqs. (\ref{integralone}) and (\ref{nontrivialint}) in
eq. (\ref{insertiontwo}), and the result so obtained in
eq. (\ref{insertone}), we have that the contribution to the
right-hand side of eq. (\ref{regbreaking}) which contains
$\epsilon^{\m_1\m_2\m_3\m_4}$ reads
\begin{equation}
  {\hat{\Delta}\cdot\Gamma^{\rm reg}[V]~\bigg\vert}_{\rm eps} =
       {\cal A}^{\rm reg}_\star~,
\label{epsanom}
\end{equation}
where
\begin{equation*}
   {\cal A}^{\rm reg}_\star = - \frac{i}{24\pi^2} \idxo\> 
      \epsilon^{\m_1\m_2\m_3\m_4}\> \mathbf{Tr}\, \Lambda \star 
         \partial_{\m_1} \Big( V_{\m_2}\star\partial_{\m_3} V_{\m_4} + 
             \frac{1}{2}\> V_{\m_2}\star V_{\m_3}\star V_{\m_4}\Big) 
   + O(\varepsilon)~.
\end{equation*}
Hence, if $\Gamma^{\rm reg}_{\rm eps}[V]$ denotes the contribution to
the regularized effective action $\Gamma^{\rm reg}[V]$ involving
$\epsilon^{\m_1\m_2\m_3\m_4}$, eqs. (\ref{regbreaking}) and
(\ref{epsanom}) imply
\begin{equation}
s_\star\Gamma^{\rm reg}_{\rm eps}[V] = {\hat{\Delta} \cdot 
    \Gamma^{\rm reg}[V]~\bigg\vert}_{\rm eps} 
          =  {\cal A}^{\rm reg}_\star~.
\label{epsbreak}
\end{equation}
It is not difficult to show that the pole part of $\Gamma^{\rm
reg}[V]$ at $\varepsilon=0$ does not depend on
$\epsilon^{\m_1\m_2\m_3\m_4}$. This, together with the observation
that any vector-like contribution to the regularized effective action
--{\it i.e} not involving $\epsilon^{\m_1\m_2\m_3\m_4}$-- can be
regularized in a gauge invariant way within the framework of
dimensional regularization, implies that it is always possible to
define a renormalized effective action
\begin{equation*}
   \Gamma^{\rm ren}[V] = \Gamma^{\rm ren}_{\rm vec-like}[V] 
       + \Gamma^{\rm ren}_{\rm eps}[V]
\end{equation*}
such that
\begin{equation*}
   s_{\star}\Gamma^{\rm ren}_{\rm vec-like}[V] = 0
\end{equation*}
and
\begin{equation*}
   s_\star \Gamma^{\rm ren}_{\rm eps}[V] = \lim_{\varepsilon\to 0}\>
         {\cal A}^{\rm reg)}_\star~ = {\cal A}_\star ~,
\end{equation*}
with ${\cal A}_\star$ as in eq. (\ref{nonsemianom-2}).  Hence
eqs. (\ref{nonsemianom-1}) and (\ref{nonsemianom-2}) follow.

Using finally that $s_\star V_\m=s V_\m$ and that $V_\m$ is a function
of $v_\m$ and $\theta^{\m\n}$ we conclude that
\begin{equation}
   s \Gamma^{\rm ren}[v;\theta] = {\cal A}_\star~.
\label{aim}
\end{equation}
This equation gives a simple expression for the anomaly if written in
terms of the noncommutative fields $V_\m$ and $\lambda$. In fact,
${\cal A}^{\rm Bardeen}_{\rm nc}$ in eq. (\ref{nonsemianom-2}) is
nothing but the noncommutative counterpart of Bardeen's ordinary
anomaly. However, in terms of the fields $v_\m$ and $\lambda$, the
anomaly is a complicated power series in $\theta^{\m\n}$ with
coefficients depending on such fields. The first term of such series
is the standard Bardeen anomaly ${\cal A}^{\rm Bardeen}$ of ordinary
spacetime,
\begin{eqnarray}
   \hskip -15pt & {\cal A}_\star = {\cal A}^{\rm Bardeen} + O(\theta) &
                                       \label{anom-series}\\[6pt]
   \hskip -15pt & {\ds {\cal A}^{\rm Bardeen} 
          = {{\cal A}_\star\big\vert}_{\,\theta=0} = 
       -\frac{i}{24\pi^2} \idx~ \epsilon^{\m_1\m_2\m_3\m_4}\> 
         \mathbf{Tr}\> \lambda \,\partial_{\m_1}\Big( 
            v_{\m_2} \partial_{\m_3} v_{\mu_4} 
               + \frac{1}{2}\> v_{\m_2} v_{\m_3} v_{\m_4}\Big).} & 
\label{anom-zeroth}
\end{eqnarray}

\section{BRS triviality of $\mathbf{\theta}\!$-dependent contributions}

The functional ${\cal A}_\star$ in eqs. (\ref{nonsemianom-1}) and
(\ref{nonsemianom-2}) has been found by explicitly computing to all
orders in $\theta^{\m\n}$ the $\epsilon^{\m_1\m_2\m_3\m_4}\!$ part of
the one-loop radiative corrections to all the 1PI Green functions of
the field $V_\m$. As is well known, only radiative corrections which
are cohomologically nontrivial with respect to the ordinary chiral
BRS operator $s$, that is to say, that can not be written as the $s$
of something, yield a true anomalous contribution. To find the true
anomaly, we must therefore identify in ${\cal A}_\star$ the
cohomologically nontrivial contributions with respect to $s$. This we
do next.

If in sections 2 and 3 we take as noncommutativity matrix
$t\theta^{\m\n}\!\!$, with $t$ a real parameter, we end up with a
noncommutative BRS chiral operator $s_\star^{(t\theta)}$ and an
anomaly ${\cal A}_{{\ds \star}}^{(t\theta)\!}$ whose expressions are
obtained from those in sections 2 and 3 by replacing $\theta^{\m\n}$
with $t\theta^{\m\n}\!\!$. Note that the dependence on $t$ of
$s_\star^{(t\theta)}$ is only through the Moyal product, which now is
with respect to $t\theta^{\m\n},$ but that no explicit $t$-dependence
is introduced (see ref. \cite{Cerchiai:2002ss}). In Appendix B we
prove that the logarithmic differential with respect to $t$ of ${\cal
A}_\star^{(t\theta)}$ is $s_\star^{(t\theta)}$ trivial, or in other
words, that there exists a functional ${\cal
B}[V^{(t\theta)}\!,t\theta]$ such that
\begin{equation}
 t\>\frac{d}{dt}\> {\cal A}_\star^{(t\theta)} 
      = s_\star^{(t\theta)}\, {\cal B}\,
            \big[V^{(t\theta)}\!,t\theta\big]~.
\label{key-lemma}
\end{equation}
Let us remark that we use the logarithmic derivative $t\frac{\ds
d}{\ds dt},$ and not the ordinary derivative $\frac{\ds d}{\ds dt}$ as
in refs. \cite{Barnich:2002pb,Cerchiai:2002ss}, to be able to write
everything in terms of the noncommutativity matrix $t\theta^{\m\n}$
and to avoid having to use both $\theta^{\m\n}$ and
$t\theta^{\m\n}$. Integrating eq. (\ref{key-lemma}) over $t$ from 0 to
1 and using that --by definition of the Seiberg-Witten map--
$s_\star^{(t\theta)} V_\m^{(t\theta)}\!=sV_\m^{(t\theta)}[v,t\theta]$,
we have
\begin{equation*}
   \int_0^1 dt~ \frac{d {\cal A}_\star^{(t\theta)}}{dt} 
     =  \int_0^1 \frac{dt}{t}~s_\star^{(t\theta)}\, {\cal B}
                  \big[V^{(t\theta)}\!,t\theta\big] 
     =  \int_0^1 \frac{dt}{t}~s\, {\cal B}
                   \big[V^{(t\theta)}[v,t\theta],t\theta\big] ~.
\end{equation*}
Recalling now that ${\cal A}^{(t\theta)}_\star={\cal A}_\star$ if
$t=1$ and ${\cal A}^{(t\theta)}_\star={\cal A}^{\rm Bardeen}$ if
$t=0$, and noting that the ordinary BRS chiral operator $s$ does not
depend on $t$, we obtain
\begin{equation}
  {\cal A}_\star =  {\cal A}^{\rm Bardeen}  - 
     s \int_0^1  \frac{dt}{t} ~
          {\cal B}_\star[V^{(t\theta)}\![v,t\theta],t\theta] ~.
\label{brs-trivial}
\end{equation}
Hence the functional ${\cal A}_\star$ found in
section 3 consists of two contributions: the standard Bardeen anomaly
${\cal A}^{\rm Bardeen}$ of commutative spacetime, and a contribution
--given by the second term in eq. (\ref{brs-trivial})-- which is
cohomologically trivial with respect to the ordinary chiral BRS
operator $s$.  Comparing with eq. (\ref{anom-series}), we conclude
that all contributions to ${\cal A}_\star$ of order one or higher in
$\theta^{\m\n}$ are cohomologically trivial, hence harmless, since
they can be absorbed by adding finite counterterms to the renormalized
effective action. Indeed, consider a new renormalized effective action
${\Gamma^\prime}^{\prime\,\rm ren} [v;\theta]$ defined by
\begin{equation}
    \Gamma^{\prime\,\rm ren} [v;\theta] =  \Gamma^{\rm ren} [v;\theta] 
      -  \int_0^1 \frac{dt}{t}
            ~{\cal B}\big[V^{(t\theta)}[v,t\theta],t\theta\big]~.
\label{newren}
\end{equation}
According to our discussion above, it follows that 
\begin{equation*}
    s \,\Gamma^{\prime\,\rm ren} [v;\theta] =  {\cal A}^{\rm Bardeen}~.
\end{equation*}
We thus conclude that the anomaly is $\theta^{\m\n}\!$-independent and
has Bardeen's form. 

\section{Conclusion}

In this paper we have calculated the chiral one-loop anomaly in 4-dimensional
noncommutative gauge theories with arbitrary compact gauge group
defined through the Seiberg-Witten map. Our main result is that for
all these theories the chiral anomaly is the same as for their
commutative counterparts. Hence any noncommutative chiral gauge theory
of this type is anomaly free to one-loop order if, and only if, 
its ordinary counterpart
is. This implies in particular that the anomaly cancellation
conditions for the noncommutative standard model \cite{Calmet:2001na}
and the noncommutative $SU(5)$ and $SO(10)$ models
\cite{Aschieri:2002mc} are the same as for the ordinary ones
\cite{Gross:pv}. We would like to emphasize that we have not found
anomaly candidates but actually computed the anomaly, since we have
calculated the relevant Feynman diagrams that produce the anomaly.

There is one key ingredient in our proof, namely that counterterms
with mass dimension greater than four should be allowed in the
renormalized effective action. This is necessary to cancel radiative
corrections which, on the one hand, do not satisfy the equation
$s\Gamma^{\rm ren}=0$ but, on the other, are cohomologically trivial
with respect to $s$. This indicates that the proper framework for
these theories is the effective field theory formalism, a proposal
that has already been made by a number of authors
\cite{Aschieri:2002mc,Wulkenhaar:2001sq,Martin:2002nr}.  If one
insists on power-counting renormalizability, then the ``safe''
representations and the safe ``groups'' of ordinary gauge theories
\cite{Georgi:1972bb} are totally unsafe for noncommutative gauge
theories, since they lead to anomalous theories \cite{Martin:2002nr}.

\section*{Acknowledgments}

CPM and FRR are grateful to CICyT, Spain for
partial support through grant No. BFM2002-00950.

\renewcommand{\thesection}{\Alph{section}}
\setcounter{section}{0}

\section{Appendix: Useful integrals}
\renewcommand{\theequation}{A.\arabic{equation}}

To obtain the $\epsilon^{\m_1\m_2\m_3\m_4}$ contribution to the
$n\!$-point functions $\Gamma^{\m_1\ldots\m_n}(p,p_1,\ldots,p_n\mid
\hat{\Delta})$ with one evanescent insertion $\hat{\Delta}$ given in
eqs. (\ref{nontrivialint}) the following integrals are needed:
\begin{equation*}
\begin{array}{l}
   {\ds \idqo ~ \frac{\hat{q}^2}{q^2\,(q-q_1)^2\,(q-p_2)^2} = 
      - \frac{1}{2}\> \frac{i}{16\pi^2} + O(\varepsilon) } \\[18pt]
   {\ds \idqo ~ \frac{\hat{q}^2 \,\bar{q}_\m} 
                     {q^2\,(q-q_1)^2\,(q-q_2)^2} =
      - \frac{1}{6}\> \frac{i}{16\pi^2}\> (\bar{q}_1 + \bar{q}_2)_\m 
      + O(\varepsilon) } \\[18pt]
   {\ds \idqo ~ \frac{\hat{q}^2\,\bar{q}_{\m_1} \bar{q}_{\m_2}} 
                     {q^2\,(q-q_1)^2\,(q-q_2)^2\,(q-q_3)^2} = 
      - \frac{1}{12}\> \frac{i}{16\pi^2}\> \bar{g}_{\m_1\m_2} 
      + O(\varepsilon) } \\[18pt]
   {\ds \idqo ~ \frac{\hat{q}^2\,\bar{q}_{\m_1} 
                      \bar{q}_{\m_2}\bar{q}_{\m_3}}
       {q^2\,(q-q_1)^2\,(q-q_2)^2\,(q-q_3)^2} = \hspace{200pt}}\\[9pt]
\hphantom{\ds \hspace{100pt}}
      {\ds =\,-\frac{1}{48}\> \frac{i}{16\pi^2}\> \sum_{i=1}^3 \left( 
          \bar{g}_{\m_1\m_2}\,\bar{q}_{i\m_3} + 
          \bar{g}_{\m_1\m_3}\,\bar{q}_{i\m_2} +
          \bar{g}_{\m_2\m_3}\,\bar{q}_{i\m_1} \right) + O(\varepsilon)
      }\\[18pt]
   {\ds \idqo ~ \frac{\hat{q}^2\,\bar{q}^2\,\bar{q}_{\m_1} 
                      \bar{q}_{\m_2}}
                     {q^2\,(q-q_1)^2\,(q-q_2)^2\,(q-q_3)^2\,(q-q_4)^2}
      = - \frac{1}{16} \frac{i}{16\pi^2}\> \bar{g}_{\m_1\m_2} 
        + O(\varepsilon) ~.}
\end{array}
\end{equation*}
\vspace{3pt}
Here $\varepsilon=\omega-2$.

\section{Appendix: Proof of eq. (4.1)}
\renewcommand{\theequation}{B.\arabic{equation}}

In what follows we will use $\omega^{\a\b}$ for $t\theta^{\a\b}$,
denote the Moyal product with respect to $\omega^{\a\b}$ by $\star$
and write a small circle $\overset{\circ}{}$ for the logarithmic
differential with respect to $t$, {\it i.e.}
\begin{equation}
   \omega^{\a\b}= t\theta^{\a\b} \qquad 
   \star = \star_{\omega} \qquad
   \overset{~\,\circ}{\cal F} = t \> \frac{d{\cal F}}{dt} ~.
\label{notation}
\end{equation}
The functional ${\cal A}^{(t\theta)}_\star\!\!$, which in this
notation we write as ${\cal A}_\star$, has a piece of order zero in
$\omega^{\a\b}$ given by ${\cal A}^{\rm Bardeen}$ in
eq. (\ref{anom-zeroth}) and a piece that collects all the higher order
terms in $\omega^{\a\b}$ and which precisely gives the contributions
to $\dotAstar$. We want to prove eq. (\ref{key-lemma}), which now
takes the form
\begin{equation}
   \dotAstar = s_\star {\cal B}~.
\label{lemma-bis}
\end{equation}
Using
\begin{equation}
    f\,\overset{\circ}{\star}\, g = \frac{1}{2} \> \omega^{\a\b}\, 
     \partial_\a f\! \star  \partial_\b g
\label{stardot}
\end{equation}
and \cite{Barnich:2002pb}
\begin{equation}
   \overset{~\circ}{V}_\m = -\frac{i}{4}\>\omega^{\a\b}\> 
         \left\{V_\a\,,\,F_{\b\m}+\partial_\b V_\m\right\}_\star \qquad\quad
   \overset{\,\circ}{\Lambda} =  \frac{i}{4}\>\omega^{\a\b}\> 
      \left\{\partial_\a\Lambda,V_\b\right\}_\star ~, 
\label{Vdot}
\end{equation}
the functional $\dotAstar$ can be expanded as a sum 
\begin{equation}
  \dotAstar = \dotAthree + \dotAfour + \dotAfive+ \dotAsix~,
\label{A-number}
\end{equation}
where $\overset{~\,\circ}{\cal A}_{\star,n}$ collects all
contributions in $\overset{~\circ}{\cal A}_\star$ of degree $n$ in the
fields $\Lambda$ and $V_\m$ (see below for their explicit
expressions). In turn, the noncommutative chiral BRS operator
$s_\star$ can be written as the sum
\begin{equation}
    s_\star = s_{\star,0} + s_{\star,1} 
\label{s-decomposition}
\end{equation}
of two operators $s_{\star,0}$ and $s_{\star,1}$ whose action on the
fields $\Lambda$ and $V_\m$ is given by
\begin{alignat}{2}
    s_{\star,0} V_\m & = \partial_\m \Lambda  & \qquad
       & s_{\star,0}\Lambda = 0 \label{s0} \\[3pt]
    s_{\star,1}V_\m & = [V_\m,\Lambda]_\star & \qquad 
        & s_{\star,1}\Lambda = -\Lambda\star\Lambda~. \label{s1}
\end{alignat}
These two operators satisfy
\begin{equation*}
   s_{\star,0}^2=0 \qquad s_{\star,0}\, s_{\star,1} + s_{\star,1}\,
   s_{\star,0} = 0 
\end{equation*}
and have the important property that $s_{\star,0}$ preserves the
degree in the fields and $s_{\star,1}$ increases it by one.  From
eqs. (\ref{A-number}) and (\ref{s-decomposition}) it follows that to
prove eq. (\ref{lemma-bis}) it is sufficient to take for ${\cal B}$ an
expansion
\begin{equation*}
  {\cal B} = {\cal B}_3 + {\cal B}_4 + {\cal B}_5 + {\cal B}_6
\end{equation*}
in the number of fields and show that
\begin{align}
  \dotAthree & = s_{\star,0}{\cal B}_3 \label{ladder-3}  \\[3pt]
  \dotAfour - s_{\star,1}{\cal B}_3 & = s_{\star,0}{\cal B}_4 
                                         \label{ladder-4} \\[3pt]
  \dotAfive - s_{\star,1}{\cal B}_4 & =  s_{\star,0}{\cal B}_5 
                                         \label{ladder-5} \\[3pt]
  \dotAsix - s_{\star,1}{\cal B}_5  & = s_{\star,0}{\cal B}_6 
                                         \label{ladder-6} \\[3pt]
            s_{\star,1}{\cal B}_6 & = 0~. \label{ladder-7}
\end{align}
Hence, to prove (\ref{lemma-bis}) all we have to do is finding
functionals ${\cal B}_3,\> {\cal B}_4,\> {\cal B}_5$ and ${\cal B}_6$
satisfying the ladder equations. To do this it is convenient to use
differential forms, so let us write eqs. (\ref{s0}) and (\ref{s1}) in
terms of differential forms. Recalling that $V= V_\m dx^\m$ and using
$\{dx^\m,s_{\star,0}\} = \{dx^\m,s_{\star,1}\} = \{\Lambda,dx^\m\}
=0$, we have
\begin{alignat}{2}
     s_{\star,0} V & = -\,d\Lambda & \qquad 
        & s_{\star,0}\Lambda = 0  \label{s0diff} \\[3pt]
    s_{\star,1} V & = - \left\{ V,\Lambda\right\}_\star & \qquad
        & s_{\star,1}\Lambda  = -\Lambda\star\Lambda~. \label{s1diff}
\end{alignat}

\subsection{Computation of ${\cal B}_{\mathbf{3}}$ and ${\cal
B}_{\mathbf{4}}$}

Taking the logarithmic differential with respect to $t$ of ${\cal
A}_\star$ and using eqs. (\ref{stardot}) and (\ref{Vdot}), it is
straightforward to see that
\begin{equation*}
   \dotAthree = - \frac{i}{24\pi^2} \int \frac{i}{2}~ \om^{\a\b}~
   \mathbf{Tr}~ \partial_\a\Lambda \star \partial_\b dV \!\star dV~.
\end{equation*}
It is clear that 
\begin{equation}
   {\cal B}_3 =   - \frac{i}{24\pi^2} \int \frac{i}{2}~ \om^{\a\b}~ 
     \mathbf{Tr}~ \left[ x\, V_\a \star \partial_\b dV\!\star dV 
       -(1-x)\,  V_\a \star dV \star \partial_\b dV\right] ~,
\label{B3}
\end{equation}
with $x$ an arbitrary parameter, solves eq. (\ref{ladder-3}). Indeed,
acting with $s_{\star,0}$ on ${\cal B}_3$, integrating by parts the
derivative $\partial_\b$ in the second term in eq. (\ref{B3}) and
neglecting the integral of a divergence, we recover $\dotAthree$. Note
that eq. (\ref{B3}) provides a one-parameter family of solutions for
${\cal B}_3$. Furthermore, to ${\cal B}_3$ we can also add a term
\begin{equation*}
   \int \omega^{\a\b}\> \mathbf{Tr}~ \partial_\a V_\b \star dV \star dV
\end{equation*}
with arbitrary coefficient, since $s_{\star,0}$ acting on it
vanishes. 

Let us move now on to eq. (\ref{ladder-4}). We first calculate
$\dotAfour$ and $s_{\star,1}{\cal B}_3$. Acting with $t\,\frac{\ds
d}{\ds dt}$ on ${\cal A}_\star$, noting eqs. (\ref{stardot}) and
(\ref{Vdot}), retaining terms of order four in the fields, using the
cyclic property of the trace $\mathbf{Tr}$ and of the integral of a
Moyal product of functions to push the ghost field $\Lambda$ to the
far left, and integrating by parts whatever partial and/or exterior
derivatives act on $\Lambda$, we obtain after some lengthy algebra
that
\begin{align*}
   \dotAfour = & - \frac{i}{24\pi^2} \int \frac{i}{4} \> \om^{\a\b}\>
     \mathbf{Tr}~ \Lambda \star \Big[ \, 
          V_\a \star dV \star \partial_\b dV 
        - V_\a \star \partial_\b dV \star dV
        + V \star \partial_\a V \star \partial_\b dV    \\ 
     &  - V \star \partial_\a dV \star \partial_\b V     
        - \partial_\a V_\b \star dV \star dV
        - dV \star dV \star \partial_\a V_\b        
        - 2\,\partial_\a V \star dV_\b \star dV         \\[6pt]
     &  + 2\, dV_\a \star dV_\b \star dV 
        - 2\, dV_\a \star \partial_\b V \star dV          
        - 2\, dV \star \partial_a V \star dV_\b 
        + 2\, dV \star dV_\a \star dV_\b                   \\[6pt]
     &  - 2\, dV \star dV_\a \star \partial_\b V
        + dV \star \partial_\a V \star \partial_\b V
        - \partial_\a V \star \partial_\b V \star dV      
        - \partial_\a V \star dV \star \partial_\b V      \\[6pt]
     &  - \partial_\a dV \star dV \star V_\b
        + 2\, \partial_\a dV \star V_\b \star dV          
        + dV \star \partial_\a dV \star V_\b          
        - 2\, dV \star V_\a \star \partial_\b dV          \\[3pt]
     &  + \partial_\a dV \star \partial_\b V \star V      
        - \partial_\a V \star \partial_\b dV \star V
        + \partial_\a dV \star V \star \partial_\b V   
        + \partial_\a V \star V \star \partial_\b dV \,\Big]
\end{align*}
Proceeding similarly for $s_{\star,1}{\cal B}_3$, and taking for
simplicity $x=1$, we have
\begin{align*}
   s_{\star,1}{\cal B}_3  & = \frac{i}{24\pi^2} \int \frac{i}{2} \> 
      \om^{\a\b}\> \mathbf{Tr}~ \Lambda \star \Big[ \, 
          V_\a \star \partial_\b dV \star dV
        + V \star dV_\a \star \partial_\b dV       \\
      & + V \star dV \star \partial_\a dV_\b    
        + V \star \partial_\a dV \star dV_\a 
        + dV_\a \star \partial_\b dV \star V       \\[3pt] 
      & + dV \star \partial_\a dV_\b \star V
        + \partial_\a dV \star dV \star V_\b 
        + \partial_\a dV \star dV_\b \star V  \,\Big]
\end{align*}
To simplify these expressions we introduce the notation
\begin{equation*}
     A^i = - \frac{i}{24\pi^2} \int \frac{i}{4}\> \om^{\a\b}~
    \mathbf{Tr}~ \Lambda \star a^i_{\a\b} ~,
\end{equation*}
with $a^i_{\a\b}$ as in Table 1. Note that $\omega^{\a\b} a^i_{\a\b}$
is a 4-form with one explicit $\omega^{\a\b}$, three explicit
derivatives and three noncommutative gauge fields. By ``explicit''
here we mean $\omega\,'{\rm s}$ and $\partial\,'{\rm
s}$ that are not hidden in the $\star\!$-product.
\begin{table}[ht]
\vspace{9pt}
\begin{center}
\renewcommand\arraystretch{1.5}\setlength{\fboxsep}{0pt}
\begin{tabular}{||l|l||l|l||l|l||}
\hline
$a^1_{\a\b}$    & ~$dV \star \partial_\a V \star \partial_\b V$  &
$a^{14}_{\a\b}$ & ~$\partial_\a V \star dV \star dV_\b$ &
$a^{27}_{\a\b}$ & ~$V \star dV_\a \star \partial_\b dV$    \\
$a^{2}_{\a\b}$  & ~$\partial_\a V \star dV \star \partial_\b V$   &
$a^{15}_{\a\b}$ & ~$\partial_\a V \star dV_\b \star dV$   &
$a^{28}_{\a\b}$ & ~$\partial_\a dV \star dV_\b \star V$  \\ 
$a^{3}_{\a\b}$  & ~$\partial_\a V \star \partial_\b V \star dV$   &
$a^{16}_{\a\b}$ & ~$dV \star dV_\a \star \partial_\b V$   &
$a^{29}_{\a\b}$ & ~$dV_\a \star \partial_\b dV \star V$   \\ 
$a^{4}_{\a\b}$  & ~$\partial_\a dV \star V \star \partial_\b V$   &
$a^{17}_{\a\b}$ & ~$dV_\a \star dV \star \partial_\b V$   &
$a^{30}_{\a\b}$ & ~$dV_\a \star V \star \partial_\b dV $   \\ 
$a^5_{\a\b}$    & ~$V \star \partial_\a dV \star \partial_\b V$   &
$a^{18}_{\a\b}$ & ~$dV_\a \star \partial_\b V \star dV$   &
$a^{31}_{\a\b}$ & ~$\partial_\a dV_\b \star dV \star V$   \\ 
$a^6_{\a\b}$    & ~$V \star \partial_\a V \star \partial_\b dV $   &
$a^{19}_{\a\b}$ & ~$\partial_\a dV \star dV \star V_\b$   &
$a^{32}_{\a\b}$ & ~$dV \star \partial_\a dV_\b \star V$   \\ 
$a^7_{\a\b}$    & ~$\partial_\a dV \star \partial_\b V \star V$   &
$a^{20}_{\a\b}$ & ~$dV \star \partial_\a dV \star V_\b$   &
$a^{33}_{\a\b}$ & ~$dV \star V \star \partial_\a dV_\b$   \\ 
$a^8_{\a\b}$    & ~$\partial_\a V \star \partial_\b dV \star V$   &
$a^{21}_{\a\b}$ & ~$dV \star V_\a \star \partial_\b dV $   &
$a^{34}_{\a\b}$ & ~$\partial_\a dV_\b \star V \star dV$   \\ 
$a^9_{\a\b}$    & ~$\partial_\a V \star V \star \partial_\b dV$   &
$a^{22}_{\a\b}$ & ~$\partial_\a dV \star V_\b \star dV$   &
$a^{35}_{\a\b}$ & ~$V \star \partial_\a dV_\b \star dV $   \\ 
$a^{10}_{\a\b}$ & ~$dV \star dV \star \partial_\a V_\b$   &
$a^{23}_{\a\b}$ & ~$V_\a \star \partial_\b dV \star dV$   &
$a^{36}_{\a\b}$ & ~$V \star dV \star \partial_\a dV_\b  $   \\ 
$a^{11}_{\a\b}$ & ~$dV \star \partial_\a V_\b \star dV$   &
$a^{24}_{\a\b}$ & ~$V_\a \star dV \star \partial_\b dV $   &
$a^{37}_{\a\b}$ & ~$dV \star dV_\a \star dV_\b$   \\ 
$a^{12}_{\a\b}$ & ~$\partial_\a V_\b \star dV \star dV$   &
$a^{25}_{\a\b}$ & ~$\partial_\a dV \star V \star dV_\b$   &
$a^{38}_{\a\b}$ & ~$dV_\a \star dV \star dV_\b $   \\ 
$a^{13}_{\a\b}$ & ~$dV \star \partial_\a V \star dV_\b $   &
$a^{26}_{\a\b}$ & ~$V \star \partial_\a dV \star dV_\b $   &
$a^{39}_{\a\b}$ & ~$dV_\a \star dV_\b \star dV$   \\ \hline
\end{tabular}
\\[12pt]
{\small \sl
Table 1: All 4-forms with three derivatives and three gauge fields.}
\end{center}
\vspace{-12pt}
\end{table}
In Table 1 we have listed all such forms that can be constructed. With
this notation $\dotAfour - s_{\star,1}{\cal B}$ reads
\begin{align}
   \dotAfour -  s_{\star,1}{\cal B} & = A^1 - A^2 + A^3 + A^4 - A^5 
         + A^6 + A^7 - A^8 + A^9 - A^{10} - A^{12}  \nonumber\\
      & - 2 A^{13} - 2 A^{15} - 2 A^{16}  - 2 A^{18} + A^{19} 
        + A^{20} - 2 A^{21} + 2 A^{22} + A^{23} \label{Afour-s1B3}\\ 
      & + A^{24} + 2 A^{26} + 2 A^{27} + 2 A^{28} + 2 A^{29} 
        + 2 A^{32} + 2 A^{36} + 2 A^{37} + 2 A^{39}\,. \nonumber 
\end{align}

Now, not all the 4-forms $w^{\a\b}a_{\a\b}^i$ in Table 1 are linearly
independent. To see this, consider {\it e.g.} the 5-form $\Omega^\b =
\omega^{\a\b}\partial_\a V\star dV \star dV\,$ and act on it with the
inner contraction
\begin{equation*}
   i_\b\equiv i_{\partial_\b}= \frac{\partial}{\partial(dx^\b)}~.
\end{equation*}
Being a 5-form in four dimensions, $\Omega^\b$ is identically zero,
and so is $i_\b$ acting on it. Hence
\begin{align*}
   0 & = i_\b\, ( \omega^{\a\b}\>\partial_\a V\star dV \star dV) \\
     & = \omega^{\a\b}\, \big[\,\partial_\a V_\b \star dV \star dV -
     \partial_\a V \star (\partial_\b V - dV_\b) \star dV -
     \partial_\a V \star dV \star (\partial_\b V - dV_\b)\,\big] \\ 
     & = \omega^{\a\b} \big(\, a^{12}_{\a\b} - a^3_{\a\b} 
      + a^{15}_{\a\b} - a^2_{\a\b} + a^{14}_{\a\b} \big) ~,
\end{align*}
which implies the relation 
\begin{equation*}
       A^{12} - A^3 + A^{15} - A^2 + A^{14} =0~.
\end{equation*}
This suggests that, to generate all the linear relations among the
functionals $A^i,$ it is enough to act with $i_\b$ on all the 5-forms
$\Omega^\b$ with one explicit $\omega^{\a\b}$, three explicit
derivatives and three noncommutative gauge fields. In listing the
forms $\Omega^\b$, two restrictions should be observed. The first one
is that it is only necessary to consider 5-forms $\Omega^\b$ with at
most two explicit derivatives acting on the same field, since in Table
1 there is no $a^i_{\a\b}$ with more than two explicit derivatives on
the same gauge field. The second one is that whenever two explicit
derivatives act on the same gauge field, they should not be both
exterior derivatives. The reason for this is that 5-forms $\Omega^\b$
containing an explicit $d^2$ do not provide, upon acting on them with
$i_\b$, any relation among the $\omega^{\a\b} a^i_{\a\b}.$ Indeed,
since $i_\b d^2 = \partial_\b d - d\partial_\b$, the action of $i_\b$
on a 5-form containing an explicit $d^2$ yields a linear combination
\begin{equation*}
    i_\b \,({\rm 5\!\!-\!\!form~with~} d^2)^\b 
    = {\rm 4\!\!-\!\!forms~with~} d^2  
    + {\rm 4\!\!-\!\!form~with~} \partial_\b d  - d \partial_\b 
\end{equation*}
of 4-forms each of which is identically zero. There are twelve
different forms $\Omega^\a$ that can be constructed satisfying these
restrictions on the derivatives, namely $\Omega^\b=\omega^{\a\b}
\tilde{a}_\a$, with $\tilde{a}_\a$ given by
\begin{alignat*}{3}
     & \partial_\a V \star dV \star dV 
     & \qquad dV_\a \star dV \star dV
     & \qquad \partial_\a dV \star dV \star V
     & \qquad \partial_\a dV \star V \star dV  & \\
     & dV \star \partial_\a V \star dV
     & \qquad dV \star dV_\a \star dV
     & \qquad dV \star \partial_\a dV \star V
     & \qquad V \star \partial_\a dV  \star dV  & \\
     & dV \star dV \star \partial_\a V 
     & \qquad dV \star dV \star dV_\a 
     & \qquad dV \star V \star \partial_\a dV 
     & \qquad V \star dV \star \partial_\a dV& ~. 
\end{alignat*}
If we act with $i_\b$ on these twelve 5-forms, we obtain the linear
relations
\begin{align*}
    A^2 + A^3 - A^{12} - A^{14} - A^{15} = 0 &    
    \qquad\quad   A^7 + A^{19} - A^{28} - A^{31} = 0  \\
    A^1 + A^3 - A^{11} - A^{13} - A^{18} = 0 &      
    \qquad\quad   A^8 - A^{20} - A^{29} + A^{32} = 0 \\
    A^1 + A^2 - A^{10} - A^{16} - A^{17} = 0 &      
    \qquad\quad  A^9 + A^{21} - A^{30} - A^{33} = 0  \\ 
    A^{12} + A^{17} + A^{18} - A^{38} - A^{39} = 0 &  
    \qquad\quad  A^4 - A^{22} - A^{25} + A^{34} = 0  \\ 
    A^{11} + A^{15} + A^{16} - A^{37} - A^{39} = 0 & 
    \qquad\quad  A^5 + A^{23} - A^{26} - A^{35} = 0  \\
    A^{10} + A^{13} + A^{14} - A^{37} - A^{38} = 0 &  
    \qquad\quad   A^6 - A^{24} - A^{27} + A^{36} = 0 ~.
\end{align*}
Solving this system of equations for $A^i$ $(i=1,\ldots, 12)$ and
substituting the solution in eq. (\ref{Afour-s1B3}), we write
$\dotAfour-s_{\star,1}{\cal B}_3$ in terms of the functionals $A^i$
$(i=13,\ldots,39)$, the result being
\begin{align}
   \dotAfour-s_{\star,1}{\cal B}_3  & = A^{13} + A^{14} - 4A^{15} 
      - 4A^{16} + A^{17} + A^{18} - 3A^{21} + 3A^{22} \nonumber \\ 
    & + 2A^{23} + 2A^{24} + A^{25} + A^{26} + 3A^{27} + 3A^{28} + A^{29} 
      + A^{30}  \label{lhs} \\
    & + A^{31} + 3A^{32} + A^{33} - A^{34} - A^{35} + A^{36} + 2A^{37} 
    - 3A^{38} + 2A^{39} ~. \nonumber 
\label{lhs}
\end{align}

We have thus obtained the left-hand side of eq. (\ref{ladder-4}) in
terms of linearly independent functionals $A^i$ $(i=13,\ldots,39)$,
each of which has one explicit $\omega^{\a\b}$ and three explicit
derivatives and has degree three in the noncommutative gauge field. It
then follows that, for eq. (\ref{ladder-4}) to have a solution, ${\cal
B}_4$ on the right-hand side must be a linear combination of
functionals
\begin{equation}
    B^r = - \frac{i}{24\pi^2} \int \frac{i}{4}\> \om^{\a\b}~
    \mathbf{Tr}~ b^r_{\a\b} ~,
\label{Bs}
\end{equation}
with $b^r_{\a\b}$ a 4-form of order two in explicit derivatives and
four in the noncommutative gauge field. With some patience, it can be
seen that there are forty such functionals $B^r$ whose $s_{\star,0}$
variation is not zero. Thirty of them can be
\begin{table}[ht]
\vspace{9pt}
\begin{center}
\begin{tabular}{||l|l||l|l||}
\hline
$b^1_{\a\b}$ & ~$V_\a \star V_\b \star dV \star dV $  &
$b^6_{\a\b}$ & ~$\partial_\a V_\b \star V \star dV \star V$    \\
$b^2_{\a\b}$ & ~$dV_\a \star V_\b \star dV \star V$  &  
$b^7_{\a\b}$ & ~$\partial_\a V_\b \star dV \star V \star V$    \\
$b^3_{\a\b}$ & ~$dV \star dV_\a \star V \star V_\b$  &
$b^8_{\a\b}$ & ~$V_\a \star dV \star \partial_\b V \star V$   \\ 
$b^4_{\a\b}$ & ~$dV_\a \star V_\b \star V \star dV$   &
$b^9_{\a\b}$ & ~$V_\a \star \partial_\b V \star V \star dV$    \\
$b^5_{\a\b}$ & ~$\partial_\a V_\b \star V \star V \star dV$  & 
$b^{10}_{\a\b}$ & ~$V \star dV \star \partial_\a V \star V_\b$  \\ 
\hline 
\end{tabular}
\\[12pt]
{\small \sl
Table 2: All 4-forms with two derivatives and four gauge fields.}
\end{center}
\vspace{-12pt}
\end{table}
written as linear combinations of the functionals $B^r$ whose
$b^r_{\a\b}$ are collected in Table 2. To illustrate that this is
indeed so, let us consider as an example
\begin{equation*}
   B = - \frac{i}{24\pi^2} \int \frac{i}{4}\> \om^{\a\b}~
       \mathbf{Tr}~ b_{\a\b} \qquad\quad 
   b_{\a\b}= V\star V_\a \star \partial_\b V \star dV~.
\end{equation*}
Clearly, this $b_{\a\b}$ in not in Table 2. However, using that
\begin{itemize}
\item[(a)] both $\mathbf{Tr}$ and the integral of a Moyal product
  of functions are cyclic,
\item[(b)] that $\partial_\a=\{i_\a,d\}$, and
\item[(c)] that $i_\b (dV)\, \star dV \star V \star V_\a =
   -dV\star i_\b ( dV \star V \star V_\a),$
\end{itemize}
and integrating by parts and neglecting total derivatives, we have
\begin{align*}
    B & \overset{\rm (a,b)}{=} \frac{i}{24\pi^2} \int \frac{i}{4}\> 
        \om^{\a\b}~ \mathbf{Tr}~ (i_\b d+ d i_\b) V \star dV
        \star V \star V_\a\\
      & \overset{\rm (c,d)}{=} - \frac{i}{24\pi^2} \int \frac{i}{4}\> 
        \om^{\a\b}~ \mathbf{Tr}~[\, dV\star i_\b\,(dV\star V \star
        V_\a) + i_\b V\star d\,(dV \star V \star V_\a)\,] \\
      & \overset{\rm ~(a)~}{=} -  \frac{i}{24\pi^2} \int \frac{i}{4}\> 
        \om^{\a\b}~ \mathbf{Tr}~( \,b^8_{\a\b} + b^3_{\a\b}
        + b^2_{\a\b}\,)\\
      &\overset{~~~~\,}{=} B^8 +B^3 +B^2 ~.
\end{align*}
Similarly, any other functional $B$ whose $b_{\a\b}$ is not in Table 2
can be expressed as a linear combination of functionals $B^r$ with
$b^r_{\a\b}$ in Table 2. It then follows that it is enough to write
for ${\cal B}_4$ 
\begin{equation}
    {\cal B}_4 = \sum_{r=1}^{10} c_r\,B^r~. 
\label{trial}
\end{equation}
To solve eq. (\ref{ladder-4}) we need the $s_{\star,0}$ variation of
${\cal B}_4$. Acting with $s_{\star,0}$ on (\ref{trial}) and writing
the result in terms of the linearly independent functionals $A^i$,
corresponding to $i=13,\ldots,39$, we obtain
\begin{align}
    s_{\star,0} {\cal B}_4 & = (-c_1 -c_2 +c_5 +c_6 + 2 c_9)\> A^{13}
     \nonumber \\
    & + (- c_1 -c_4 + c_5 + c_6 + c_8 + 3 c_9)\> A^{14}  \nonumber \\
    & + (- 2c_3 - c_5 + c_7 + c_8 + c_{10})\> (A^{15} + A^{16})
    \nonumber \\
    & + (c_1 - 2 c_2 - c_4 - c_6 - c_7 +2 c_8 - c_9 +c_{10})\>A^{17}
    \nonumber \\  
    & + (c_1-c_2 - 2 c_4 - c_6 - c_7 + c_8 + c_{10})\> A^{18}
      \nonumber \\ 
    & + ( c_1 + c_8 - c_9)\>  (A^{19} + A^{20}) \nonumber \\
    & + (c_3-c_8)\> ( -A^{21} + A^{22} + A^{27} + A^{28}) \nonumber \\ 
    & + (c_1 -c_2 - c_4)\> A^{23}  \nonumber \\
    & + (c_1 -c_2 - c_4+ c_8 + c_{10})\>  A^{24} \nonumber \\
    & - c_2\, (A^{25} + A^{29}) \nonumber \\
    & + (-c_4+c_9 + c_{10})\> (A^{26} + A^{30}) \nonumber \\
    & + ( c_2+ c_7-c_8)\> A^{31} \nonumber \\ 
    & + (c_3+c_5)\> A^{32} \nonumber \\
    & + (c_2-c_6-c_9)\> A^{33} \nonumber \\
    & + (c_4+c_6+c_{10})\> A^{34} \nonumber \\
    & + ( c_3-c_7-c_8-c_{10}) \> A^{35} \nonumber \\
    & + ( c_4-c_5-c_9)\> A^{36}  \nonumber \\
    & + ( c_1+c_3-c_4-c_6-c_7-c_9)\>  A^{37} \nonumber \\
    & + (c_2-c_3+c_4-c_5+c_7-c_9)  \> A^{38} \nonumber \\
    & + (-c_1+c_34+c_4+c_5+c_6+c_8-c_9+c_{10})\> A^{39}~. \label{rhs}
\end{align}
Substituting now eqs. (\ref{lhs}) and (\ref{rhs}) in eq. (\ref{ladder-4})
and equating the coefficients of $A^i$ $(i=13,\ldots,39)$ on both
sides, we obtain a system of 21 equations with unknowns
$c_1,\ldots,c_{10}.$ Its solution is 
\begin{alignat*}{5}
  c_1&=y+z  \qquad &c_2 &=-1 &c_3&=3-z \qquad  &c_4&=-1 +y+z 
      \qquad  &c_5&=z  \\[3pt]
  c_6&=-y     &c_7&=2-z \qquad  &c_8&=-z  &c_9&=y  &c_{10}&=z~,
\end{alignat*}
where $y$ and $z$ are arbitrary parameters. This provides a
two-parameter family of functionals ${\cal B}_4$ for which
eq. (\ref{ladder-4}) holds. Note that if we take $y=z=0$, then ${\cal
B}_4$ only has four terms. 

\subsection{Calculation of ${\cal B}_{\mathbf{5}}$ and 
${\cal B}_{\mathbf{6}}$}

One may proceed analogously as for ${\cal B}_4$ and explicitly compute
${\cal B}_5$ and ${\cal B}_6$. Here, instead, we present an
alternative method which uses cohomological techniques. To apply them
we shall employ the approach of ref. \cite{Barnich:2002pb} which
introduces gauge fields $v_\mu^A$ and ghost fields $\lambda^A$ not
only for the Lie algebra $\lieg$ of the gauge group $G$ but also for
the whole enveloping algebra $\cU= \{T_A\}= \{T_\LA,T_i\}$ in which
$V_\mu$ and $\Lambda$ take values.  Here the index $\LA$ runs over the
elements of $\lieg$, so that in the notation of section 2 one has
$\{T_\LA\}= \{(T^k)^a,T^l\}$, while the index $i$ runs over the
complementary elements of $\cU$. As shown in
ref. \cite{Barnich:2002pb}, the standard Seiberg-Witten map can be
extended to include $\cU$-valued fields $v_\mu$ and $\lambda$
satisfying
\begin{align}
    s v^A_\m &=\6_\m\la^A + f_{BC}{}^A\, v^B_\m \la^C 
                             \label{BRSext1} \\[3pt]
    s \la^A &= \frac 12\> \la^B \la^C f_{CB}{}^A ~, \label{BRSext2}
\end{align}
with $f_{AB}{}^C$ the structure constants of the Lie algebra $\cU$,
given by $[T_A,T_B]=\f ABC T_C$. Of course, $\lieg$ being a subalgebra
of ${\cal U}$ means $\f \LA\LB{i}=0$ and implies that the BRS
transformations above are subject to the truncation conditions
\begin{eqnarray}
 {sv_\mu^A\,\Big\vert}_{v_\mu^i=\lambda^i=0} &\!\!=\!\!& 
        \left\{ \begin{array}{ll} 
                \6_\mu \la^\LA+\f \LB\LC\LA\, v_\mu^\LB \lambda^\LC 
                                 & \mbox{if $A=\LA$}\\[4pt]
                \f \LB\LC{i}\, v_\mu^\LB \lambda^\LC=0 & \mbox{if $A=i$}
                \end{array} 
        \right. \label{truncation1} \\[6pt]
{s\lambda^A\,\Big\vert}_{v_\mu^i=\lambda^i=0} &\!\!=\!\!& 
        \left\{ \begin{array}{ll}
             \frac 12\, \la^\LB\lambda^\LC\,\f \LB\LC\LA  
                                 & ~~~\mbox{if $A=\LA$}\\[4pt]
         \frac 12\,\la^\LB\la^\LC\,\f \LC\LB{i}=0 &~~~\mbox{if $A=i\,.$}
                \end{array} 
        \right. \label{truncation2}
\end{eqnarray} 
The extended Seiberg-Witten map is defined by demanding 
\begin{equation}
   s_\star V_\m^A=sV_\m^A \qquad s_\star \Lambda^A= s\Lambda^A~,
\label{SWequationsext}
\end{equation}
subject to the usual boundary conditions and with $s_\star$ defined by
\begin{equation}
   s_\star V^A_\m= \6_\m \Lambda^A + f_{BC}{}^A\, V^B_\m \Lambda^C
   \qquad 
   s_\star\Lambda^A= \frac{1}{2}\> \Lambda^B\Lambda^C\,f_{CB}{}^A~.
\label{ncBRSext}
\end{equation}
By setting in it all fields $v_\mu^i$ and $\lambda^i$ to zero, the
standard Seiberg-Witten map is recovered. Furthermore, the truncation
conditions imply that all formulas that hold for ${\cal U}\!$-valued
fields $v^A_\m$ and $\la^A$ will also hold for $\lieg\!$-valued fields
$v^{\bar{a}}_\m$ and $\la^{\bar{a}}$, and in particular
eq. (\ref{lemma-bis}) that we want to prove. The idea is then to
demonstrate eq. (\ref{lemma-bis}) for the extended Seiberg-Witten map.

We start from the fact that $\cA_\star$ satisfies the anomaly
consistency condition $s_\star\cA_\star=0$ which follows from eq.
(\ref{lemma-bis}) because of $s_\star^2=0$.  In terms of the
commutative fields $v_\mu^A$ and $\lambda^A$, one has $s\cA_\star=0$.
This implies $s\dotAstar=0$ since $s$ commutes with the logarithmic
derivative with respect to $\defpar$.  Using (\ref{SWequationsext})
again, one concludes $s_\star\dotAstar=0$ which decomposes into
\begin{align}
  s_{\star,0} \dotAthree&=0 \label{E5} \\[3pt]
  s_{\star,0} \dotAfour+s_{\star,1} \dotAthree &=0 \label{E6} \\[3pt] 
  s_{\star,0} \dotAfive+s_{\star,1} \dotAfour &=0 \label{E7} \\[3pt]
  s_{\star,0} \dotAsix+s_{\star,1} \dotAfive&=0 \label{E8} \\[3pt]
  s_{\star,1}\dotAsix&=0~.\label{E9}
\end{align}
In the previous subsection we have shown by explicit computation that
(\ref{E5}) and (\ref{E6}) imply $\,\dotAthree\!=s_{\star,0}\cB_3\,$
and $\,\dotAfour\!=s_{\star,0}\cB_4+s_{\star,1}\cB_3$. We shall now
show by cohomological means that the remaining equations imply
$\dotAfive\!=s_{\star,0}\cB_5+s_{\star,1}\cB_4$,
$\,\dotAsix\!=s_{\star,0}\cB_6+s_{\star,1}\cB_5\,$ and
$\,s_{\star,1}\cB_6=0$, which will complete the proof of equations
(\ref{ladder-3}) to (\ref{ladder-7}).

To that end we first derive a result on the cohomology of
$s_{\star,0}$ in the space $\cF_\ST$ of integrated
$\star\!$-polynomials in the fields $V_\mu^A$, $\Lambda^A$ and their
derivatives. An element of this space is a linear combination, with
coefficients that may depend on $\omega^{\alpha\beta}$, of terms of
the form
\begin{equation*} 
     \int d^4x\ a_1\ST a_2 \ST \dots \ST a_n, 
\end{equation*}
with $n$ finite and each $a_i$ one of our basic variables ($V_\mu^A$,
$\Lambda^A$ and their derivatives), 
\begin{equation*}
  a_i\in \{ V_\mu^A,\Lambda^A,\6_\mu V_\nu^A,\6_\mu \Lambda^A,
            \6_\mu\6_\nu V_\rho^A,\6_\mu\6_\nu \Lambda^A,\dots  \}~. 
\end{equation*}
It is obvious why this cohomology is relevant to the present case.
Using the result $\,\dotAfour\!=s_{\star,0}\cB_4+s_{\star,1}\cB_3$
from Subappendix B.1 in eq. (\ref{E7}) and noting that
$s_{\star,1}^2=0$, we obtain $s_{\star,0} (\dotAfive-s_{\star,1}\cB_4)
= 0$, with $\,\dotAfive-s_{\star,1}\cB_4\,$ obviously in
$\cF_\ST$. Our aim is to show that this implies
$\dotAfive-s_{\star,1}\cB_4= s_{\star,0}\cB_5$ for some
$\cB_5\in\cF_\ST$, or in other words that $\dotAfive-s_{\star,1}\cB_4$
is trivial in the $s_{\star,0}\!$-cohomology in $\cF_\ST$. Assume that
we have shown this. Inserting the result in (\ref{E8}) and proceeding
similarly yields $\,s_{\star,0} (\dotAsix-s_{\star,1}\cB_5)=0$. Again,
we want to show that $\dotAsix-s_{\star,1}\cB_5=s_{\star,0}\cB_6$ for
some $\cB_6\in\cF_\ST$ and thus that $\dotAsix-s_{\star,1}\cB_5$ is
also trivial in the $s_{\star,0}\!$-cohomology in $\cF_\ST$. Note that
(\ref{ladder-3}) and (\ref{ladder-4}) actually express analogous
results, namely the triviality of $\dotAthree$ and
$\dotAfour-s_{\star,1}\cB_3$ in the same cohomology. However, as it
will become clear below, they cannot be proved by means of the result
on the cohomology for $s_{\star,0}$ in $\cF_\ST$ that we derive in the
sequel and therefore have to be shown by other methods. 

To examine the $s_{\star,0}\!$-cohomology in $\cF_\ST$ we adapt
methods developed in ref. \cite{Brandt:1989gy} for the computation of
the cohomology of $s_0$.  We first derive a result on the
$s_{\star,0}\!$-cohomology in the space $\cP_\ST$ of non-integrated
$\star\!$-polynomials. For that purpose we introduce the following
variables $u^\ell$, $v^\ell$ and $w^i$: 
\begin{align}
   \{u^\ell\} &= \{V^A_\m,\6_{(\m}V^A_{\n)},\dots,
                  \6_{(\m_1}\dots\6_{\m_k}V^A_{\m_{k+1})},\dots\} 
                                              \label{u} \\[3pt]
   \{v^\ell\} &= \{s_{\star,0}u^\ell\} 
               = \{\6_\m \Lambda^A, \6_{(\m}\6_{\n)}\Lambda^A,\dots,
                   \6_{(\m_1}\dots\6_{\m_{k+1})}\Lambda^A,\dots\}
                                              \label{v} \\[3pt]
   \{w^i\} &= \{\Lambda^A,\6_{[\m}V^A_{\n]},\dots,
                \6_{\m_1}\dots\6_{\m_k}\6_{[\m}V^A_{\n]}, \dots\}~. 
                                              \label{w}
\end{align}
Evidently every $\star\!$-polynomial in the fields $V^A_\mu$,
$\Lambda^A$ and their derivatives can be expressed as a
$\star\!$-polynomial in the variables $u^\ell$, $v^\ell$, $w^i$ and
vice versa\footnote{The set of $w$'s is actually overcomplete because
the $w$'s are not all linearly independent owing to the identities
$\6_{[\mu}\6_{\nu}V_{\rho]}=0$ and their derivatives. However this
does not matter to our arguments.}.  On non-integrated
$\star\!$-monomials in $u^\ell$, $v^\ell$, $w^i$ we define the
operation $\varrho$ through
\begin{align*}
    \varrho\, (\7a_1 &\ST \7a_2 \ST \dots \ST \7a_n) =      \\ 
    &= \frac 1n\> \Big(u^\ell\,\frac{\6\7a_1}{\6v^\ell}\Big) \ST
       \7a_2 \ST \dots \ST \7a_n                            \\ 
    &+ \frac 1n\sum_{i=2}^{n-1}
       (-)^{|\7a_1|+|\7a_2|+\ldots+|\7a_{i-1}|} 
       \7a_1 \ST \dots \ST \7a_{i-1} \ST 
       \Big(u^\ell\,\frac{\6\7a_i}{\6v^\ell}\Big) \ST \7a_{i+1} 
       \ST \dots \ST \7a_n                                  \\ 
    &+ \frac 1n\> (-)^{|\7a_1|+|\7a_2|+\ldots+|\7a_{n-1}|} 
       \7a_1 \ST \7a_2 \ST \dots \ST \7a_{n-1}\ST 
       \Big(u^\ell\,\frac{\6\7a_n}{\6v^\ell}\Big)~, 
\end{align*}
where $\7a_i$ is any of the variables $u^\ell$, $v^\ell$, $w^i$, 
\begin{equation*}
   \7a_i \in \{u^\ell, v^\ell, w^i\}~, 
\end{equation*}
and $|\7a_i|$ is the Grassmann parity of $\7a_i$, which is 0 for
$V_\mu^A$ and its derivatives, and 1 for the $\Lambda^A$ and its
derivatives.  Extending the definition of $\varrho$ by linearity from
$\star\!$-monomials to $\star\!$-polynomials, we have that the
anticommutator of $s_{\star,0}$ and $\varrho$ evaluated on an
arbitrary $\star\!$-polynomial $p_\ST(u,v,w)\in\cP_\ST$ gives the
difference 
\begin{equation}
    \{s_{\star,0},\varrho\}\,p_\ST(u,v,w)= p_\ST(u,v,w)-p_\ST(0,0,w)~,
\label{hom1}
\end{equation}
where $p_\ST(0,0,w)$ denotes the $\star\!$-polynomial that arises from
$p_\ST(u,v,w)$ by setting to zero all $u^\ell$ and $v^\ell$
{\em before} evaluating the star-products --for example, for
$p_\ST=V^A_\m\ST V^B_\n$ one has $p_\ST(0,0,w)=0$.  Applying now
eq. (\ref{hom1}) to an $s_{\star,0}$-closed $\star\!$-polynomial, {\it
i.e.} to a $p_\ST$ satisfying $s_{\star,0} p_\ST=0$, and using that
all $w^i$ are $s_{\star,0}$-closed, we obtain
\begin{equation}
   s_{\star,0}\, p_\ST(u,v,w)=0 \quad \Leftrightarrow \quad
   p_\ST(u,v,w)=p_\ST(0,0,w) + s_{\star,0}\, \varrho\, p_\ST(u,v,w)~.
\label{s0coh}
\end{equation}
In particular, an $s_{\star,0}\!$-closed $\star\!$-polynomial
$p_\ST(u,v,w)$ with $p_\ST(0,0,w)=0$ is the $s_{\star,0}\!$-variation
of the star-polynomial $\varrho\, p_\ST(u,v,w)$.

Result (\ref{s0coh}) cannot be used directly for our purposes since it
applies only to $\star\!$-polynomials but not to integrated
$\star\!$-polynomials, which is what we had initially. This makes a
difference because, by definition, an integrated $\star\!$-polynomial
is $s_{\star,0}\!$-closed when the $s_{\star,0}\!$-transformation of
its integrand is a total divergence: 
\begin{equation*}
    s_{\star,0}\, f_\ST=0 ~~{\rm with}~~ 
    f_\ST={\ds \int}\! d^4x\, p_\ST \quad \Leftrightarrow\quad 
    s_{\star,0}\,p_\ST = \6_\m\omega^\m ~~{\rm for~some~~} \omega^\m~. 
\end{equation*}
Since $\varrho$ does not commute with $\6_\mu$ we cannot directly
apply the result above to this case. To escape this problem we
consider the variational derivatives of the equation
$s_{\star,0}\,f_\ST=0$ with respect to $V_\mu^A$ and $\Lambda^A$. This
yields
\begin{equation}
   s_{\star,0}\, f_\ST=0\,, \quad f_\ST \in \cF_\ST 
       \quad  \Rightarrow \quad  
   s_{\star,0}\,\frac{\delta f_\ST}{\delta V_\m^A}=0\,, \quad 
   s_{\star,0}\,\frac{\delta f_\ST}{\delta \Lambda^A} 
      + \6_\m\,\frac{\delta f_\ST}{\delta V_\m^A} = 0~.
\label{var}
\end{equation}
It can be readily checked that the variational derivative of any
element $f_\ST\in\cF_\ST$ with respect to $V_\mu^A$ or $\Lambda^A$ is
a $\star\!$-polynomial in $\cP_\ST$.  Suppose now that $\delta
f_\ST/\delta V_\mu^A$ vanishes at $u^\ell=v^\ell=0$ in the sense
explained above. Using the first equation in (\ref{var}) and
eq. (\ref{s0coh}) we then conclude that $\delta f_\ST/\delta V_\mu^A$
is the $s_{\star,0}\!$-variation of $\varrho\,(\delta f_\ST/\delta
V_\mu^A)$:
\begin{equation*}
   \bigg[\frac{\delta f_\ST}{\delta V_\m^A}\bigg](0,0,w) = 0 
   \quad \Rightarrow\quad
   \frac{\delta f_\ST}{\delta V_\m^A} = s_{\star,0}\,\varrho\,
        \frac{\delta f_\ST}{\delta V_\m^A} ~.
\end{equation*}
Using this in the second equation in (\ref{var}) we obtain
\begin{equation}
   s_{\star,0} \left(\frac{\d f_\ST}{\d \Lambda^A} 
      + \6_\m\,\varrho\,\frac{\d f_\ST}{\d V_\m^A}\right) = ~0.
\label{var2}
\end{equation}
Applying (\ref{s0coh}) once again we conclude that the term in
parentheses is $s_{\star,0}\,\varrho(\dots)$ provided it vanishes at
$u^\ell=v^\ell=0$ in the sense above. Note that here $\varrho(\dots)$
has ghost number $\gh(f_\ST)-2$, with $\gh(f_\ST)$ the ghost number of
$f_\ST$ and $\gh(V)=0$ and $\gh(\Lambda)=1$. Since
$\star\!$-polynomials $p_\ST(u,v,w)$ have non-negative ghost numbers,
$\varrho(\dots)$ vanishes when $f_\ST$ has ghost number 1, which
is the case we are interested in. We thus conclude that
\begin{equation}
   \bigg[\,\frac{\d f_\ST}{\d \Lambda^A} + \6_\m\,\varrho\, 
      \frac{\d f_\ST}{\d V_\m^A}\bigg](0,0,w) = 0\,,
         \quad \gh(f_\ST)=1 \quad\Rightarrow\quad 
   \frac{\d f_\ST}{\d \Lambda^A} = -\6_\m\,\varrho\, 
      \frac{\delta f_\ST}{\delta V_\mu^A} ~.
\label{var3}
\end{equation}
Finally we reconstruct $f_\ST$ from its variational derivatives,
neglecting integrated divergences, using the general formula
\begin{equation}
    f_\ST[V,\Lambda] = \int\! d^4x \int_0^1 \frac{d\tau}{\tau}\,
       \left( V_\m^A \ST \frac{\d f_\ST}{\d V_\m^A}
            + \Lambda^A \ST \frac{\d f_\ST}{\d \Lambda^A} \right) 
       [\tau V,\tau \Lambda]~,
\label{reco}
\end{equation}
valid for every functional $f_\ST$. Using eqs. (\ref{var2}) and
(\ref{var3}) in (\ref{reco}) we obtain
\begin{align*}
    f_\ST[V,\Lambda] &= \int\! d^4x \int_0^1 \frac{d\tau}{\tau}\, 
      \left( V_\m^A \ST s_{\star,0}\,\varrho\, 
                                \frac{\d f_\ST}{\d V_\m^A}
           - \Lambda^A \ST \6_\m\,\varrho\, 
                                \frac{\d f_\ST}{\d V_\m^A}
      \right) [\tau V,\tau \Lambda] \\[6pt]
    & = \int\! d^4x \int_0^1 \frac{d\tau}{\tau}\, 
      \left( V_\m^A \ST s_{\star,0}\,\varrho\, 
                                    \frac{\d f_\ST}{\d V_\m^A}
           + (s_{\star,0} V_\m^A) \ST \varrho\, 
                                     \frac{\d f_\ST}{\d V_\m^A}
      \right) [\tau V,\tau \Lambda] \\[6pt]
    & = s_{\star,0} \int\! d^4x \int_0^1 \frac{d\tau}{\tau}\, 
      \left(V_\m^A \ST \varrho\,\frac{\d f_\ST}{\d V_\m^A}
      \right) [\tau V,\tau \Lambda] ~,
\end{align*}
where we have used integration by parts and $s_{\star,0}V_\m^A = \6_\m
\Lambda^A$. We have thus shown that
\begin{equation}
\begin{array}{c}
    {\ds s_{\star,0}f_\ST=0\,, \quad \gh(f_\ST)=1\,, \quad 
    \bigg[\frac{\d f_\ST}{\d V_\m^A}\bigg](0,0,w) 
          = \bigg[ \frac{\d f_\ST}{\d \Lambda^A} 
                 + \6_\m\,\varrho\,\frac{\d f_\ST}{\d V_\m^A} 
            \bigg]\, (0,0,w) = 0 } \\[15pt]
    {\ds \Rightarrow \quad f_\ST = s_{\star,0} \int\! d^4x \int_0^1
       \frac{d\tau}{\tau}\, \bigg( V_\m^A \ST \varrho\,
          \frac{\d f_\ST}{\d V_\m^A}\bigg) [\tau V,\tau \Lambda]~,}
\end{array}
\label{s0F}
\end{equation}
which is the result for the $s_{\star,0}\!$-cohomology in $\cF_\ST$ we
will use to prove eqs. (\ref{E7})-(\ref{E9}).

Consider now $\dotAfive-s_{\star,1}\cB_4$. It is an
$s_{\star,0}\!$-closed integrated $\star\!$-polynomial with ghost
number 1 whose integrand is order 5 in the fields $V_\mu^A$ and
$\Lambda^A$, has mass dimension 4 --recall that
$\Dim(V_\mu)=\Dim(\6_\mu)=1$, $\Dim(\Lambda)=0$,
$\Dim(\omega^{\alpha\beta})=-2$-- and contains one explicit
$\omega^{\a\b}$.  It follows that the integrand is a linear
combination of $\star\!$-monomials $\omega^{\alpha\beta}
a_1\ST\dots\ST a_5$, where it can be assumed that one of the $a_i$ is
an undifferentiated $\Lambda$ (for one can remove all derivatives from
$\Lambda$ using integrations by parts, if necessary) while the
remaining $a_i\,'{\rm s}$ are either of type $\{V,V,\6V,\6V\}$ or
$\{V,V,V,\6\6V\}$. It is easy to verify that this in turn implies
\begin{align}
   \bigg[\,\frac{\d (\dotAfive - s_{\star,1} \cB_4)}{\d V_\m^A} \,
   \bigg]\,(0,0,w) & = 0 \label{EC1} \\
   \bigg[\,\frac{\d (\dotAfive - s_{\star,1} \cB_4)}{\d \Lambda^A} 
        + \6_\m\,\varrho\,\frac{\d(\dotAfive - s_{\star,1}\cB_4)}
                               {\d V_\m^A}\, \bigg]\,(0,0,w) &=0 ~.
                         \label{EC2}
\end{align}
Eq. (\ref{s0F}) can then be used and yields $\,\dotAfive -
s_{\star,1}\cB_4 = s_{\star,0}\cB_5$, with 
\begin{equation*}
    \cB_5=\int\! d^4x \int_0^1 \frac{d\tau}{\tau}\> \bigg( 
      V_\m^A \ST \varrho\, \frac{\d (\dotAfive - s_{\star,1}\cB_4)} 
                                {\delta V_\mu^A} \bigg)\, 
    [\tau V,\tau \Lambda]~.  
\end{equation*}
This proves eq. (\ref{ladder-5}) for ${\cal U}\!$-valued fields, hence
for $\lieg\!$-valued fields, as we wanted to show.  
The functional $\,\dotAsix-s_{\star,1}\cB_5\,$ can be treated
analogously. Its integrand is order 6 in the fields, has mass
dimension 4, ghost number 1 and one explicit $\omega^{\a\b}$. It is
thus a linear combination of $\star\!$-monomials $\omega^{\a\b} a_1
\ST \dots \ST a_6$, where it can be assumed that the set of $a_i$ has
the structure $\{\Lambda,V,V,V,V,\6V\}$.  This makes it obvious that
$\dotAsix-s_{\star,1}\cB_5$ satisfies
\begin{align}
   \bigg[\,\frac{\d (\dotAsix-s_{\star,1}\cB_5)}{\d V_\m^A}\,
   \bigg]\, (0,0,w) &= 0 \label{EC3} \\
   \bigg[\, \frac{\d (\dotAsix-s_{\star,1}\cB_5)}{\d \Lambda^A} 
      + \6_\m\,\varrho\,\frac{\d (\dotAsix-s_{\star,1}\cB_5)}
                             {\d V_\m^A}\, \bigg]\,(0,0,w) & = 0~.
\label{EC4}
\end{align}
Eq. (\ref{s0F}) then implies $\dotAsix - s_{\star,1}\cB_5 =
s_{\star,0}\cB_6$, with
\begin{equation*}
   \cB_6 = \int\! d^4x \int_0^1 \frac{d\tau}{\tau}\, \bigg( V_\m^A \ST 
     \varrho\, \frac{\d (\dotAsix-s_{\star,1}\cB_5)}{\d V_\m^A}\,
     \bigg)\, [\tau V,\tau \Lambda]~,
\end{equation*}
which proves eq. (\ref{ladder-6}).
Finally we have to show that eq. (\ref{ladder-7}) holds.  This is very
easy. The integrand of $\cB_6$ is a $\star\!$-polynomial of order 6 in
the fields, has mass dimension 4, ghost number 0 and one explicit
$\omega^{\a\b}$.  It is thus a linear combination of
$\star\!$-monomials $\omega^{\a\b} a_1 \ST \dots \ST a_6$, where all
$a_i$ are undifferentiated $V\,'{\rm s}$. Furthermore, by
construction, it can be written as a trace $\mathbf{Tr}$. The latter
implies already $s_{\star,1}\cB_6=0$, since 
\begin{equation*}
  s_{\star,1}\,\mathbf{Tr}\,\Big( V_{\m_1}\ST\dots\ST V_{\m_6} \Big) =  
     \mathbf{Tr}\,\Big[ V_{\m_1}\ST\dots\ST V_{\m_6},\Lambda\Big]_\ST
\end{equation*}
is a divergence.

We close by remarking that eq. (\ref{s0F}) cannot be used to prove
that $\dotAthree$ and $\dotAfour-s_{\star,1}\cB_3$ are trivial in the
$s_{\star,0}\!$-cohomology in $\cF_\ST$ because the $\frac{\d}{\d
V^A_\m}$ and $\frac{\d}{\d\Lambda^A} + {\scs \6_\m \varrho\,
}\frac{\d}{\d V^A_\m}$ acting on them do not vanish at
$u^\ell=v^\ell=0$ in the sense explained above, contrary to what
happens for $\dotAfive-s_{\star,1}\cB_4$ and
$\,\dotAsix-s_{\star,1}\cB_5$ --see eqs. (\ref{EC1}), (\ref{EC2}),
(\ref{EC3}) and (\ref{EC4}).

\end{document}